\documentclass[]{sig-alternate}

\usepackage{mathtools}

\usepackage{cite}
\usepackage{paralist}
\usepackage{graphicx}

\usepackage{tikz}                                                               
\usetikzlibrary{calc, shapes, backgrounds,arrows}                               
\usepackage{subcaption}                                                         
\tikzset{                                                                       
   >=latex,shorten <=2pt,shorten >=2pt,                                          
   inner sep=1pt,                                                                
   dot/.style={->},                                                              
   dash/.style={<->,dashed},                                                     
   proc/.style={}
}          

\usetikzlibrary{calc,shapes,arrows,chains}
\usepackage{color}
\usepackage{url}

\newcommand{\subparagraph}{}
\usepackage{titlesec}
\titleformat*{\subsection}{\bfseries}

\usepackage{etoolbox}
\usepackage[hidelinks]{hyperref}
\usepackage[capitalise]{cleveref}

\newtoggle{TR}
\togglefalse{TR}
\toggletrue{TR}
\makeatletter
\def\@copyrightspace{\relax}
\makeatother

\newtheorem{definition}{Definition}
\newtheorem{theorem}{Theorem}
\newtheorem{lemma}{Lemma}
\newtheorem{corollary}{Corollary}

\DeclarePairedDelimiter\abs{\lvert}{\rvert}
\makeatletter
\let\oldabs\abs
\def\abs{\@ifstar{\oldabs}{\oldabs*}}
\makeatother

\newcommand\set[1]{\left\lbrace #1 \right\rbrace}

\newcommand{\equalityref}[1]{\hyperref[#1]{Equality~\eqref{#1}}}
\newcommand{\inequalityref}[1]{\hyperref[#1]{Inequality~\eqref{#1}}}
\newcommand{\FAES}{$\mathtt{FAES}$}
\newcommand{\ECS}{\mathtt{ECS}}

\newcommand{\MA}{message adversary}
\newcommand{\MAs}{message adversaries}
\newcommand{\MAD}{\mathtt{MA}}
\newcommand{\MAT}{\mathtt{MAT}}
\newcommand{\MAA}{\mathtt{MAA}}
\newcommand{\EStable}{\lozenge  \mathtt{STABLE}(D)}
\newcommand{\AltEStable}{\lozenge  \mathtt{STABLE}'(D)}
\newcommand{\Liveness}{\lozenge  \mathtt{STABILITY}}
\newcommand{\AltLiveness}{\lozenge  \mathtt{STABILITY}'}
\newcommand{\Safety}{\mathtt{STICKY}}
\newcommand{\AltSafety}{\mathtt{STICKY}'}

\newcommand{\rST}{r_\text{stab}}
\newcommand{\rGST}{\rST}

\newcommand{\ra}{\rightarrow}
\newcommand{\G}{\mathcal{G}}
\newcommand{\A}{\mathcal{A}}
\newcommand{\Gr}{\mathcal{G}^r}

\newcommand{\Seq}[3]{(\G^{#1})_{#1 = #2}^{#3}}
\newcommand{\Seqr}[2]{\Seq{r}{#1}{#2}}
\newcommand{\SeqInf}{\Seqr{1}{\infty}}
\newcommand{\SeqrI}{(\G^r)_{r \in I}}
\newcommand{\SeqrJ}{(\G^r)_{r \in I'}}
\newcommand\view[1]{\widehat{#1}}
\newcommand\ApproxGp[1]{\view{\G}[#1]}
\newcommand\ApproxG[3]{\view{\G}_{#2}^{#3}[#1]}
\newcommand{\Exec}[1]{\langle #1 \rangle}
\DeclareMathOperator{\roots}{roots}
\DeclareMathOperator{\CPast}{CP}

\renewcommand{\geq}{\geqslant}
\renewcommand{\leq}{\leqslant}
\renewcommand{\epsilon}{\varepsilon}
\newcommand{\state}[2]{{#1}^{#2}}

\newcommand{\CP}[3]{\CPast_{#1}^{#3}({#2})}
\newcommand{\infl}[4]{{#1} \in \CP{#3}{#2}{#4}}
\newcommand{\ninfl}[4]{{#1} \notin \CP{#3}{#2}{#4}}
\newcommand{\setinfl}[4]{{#1} \subseteq \CP{#3}{#2}{#4}}

\def\bGST{b_\text{stab}}

\def\rSR{r_\text{sr}}
\def\A{\mathcal{A}}

\newcommand{\todo}[1]{}
\renewcommand{\todo}[1]{{\color{red} TODO: {#1}}}

\usepackage{fixltx2e}
\usepackage{comment}

\numberofauthors{3}
\author{
\alignauthor
Manfred Schwarz\\
\affaddr{TU Wien}\\
\affaddr{Vienna, Austria}\\
\texttt{mschwarz@ecs.tuwien.ac.at}
\alignauthor
Kyrill Winkler\\
\affaddr{TU Wien}\\
\affaddr{Vienna, Austria}\\
\texttt{kwinkler@ecs.tuwien.ac.at}
\alignauthor
Ulrich Schmid\\
\affaddr{TU Wien}\\
\affaddr{Vienna, Austria}\\
\texttt{s@ecs.tuwien.ac.at}
}

\begin{document}

\title{Fast Consensus under Eventually Stabilizing Message Adversaries}

\maketitle
\begin{abstract}
This paper is devoted to deterministic consensus in synchronous 
dynamic networks with unidirectional links, which are under the control of 
an omniscient message adversary. Motivated by unpredictable node/system initialization 
times and long-lasting periods of massive transient faults, we consider message adversaries 
that 
guarantee periods of less erratic message loss only \emph{eventually}: We present a tight
bound of $2D+1$ for the termination time of consensus under a message adversary that
eventually guarantees a single vertex-stable root component with dynamic network
diameter $D$, as well as a simple algorithm that matches this bound. It effectively
halves the termination time $4D+1$ achieved by an existing consensus algorithm, 
which also works under our message adversary. We also introduce a generalized, 
considerably stronger variant of our message adversary, and show that 
our new algorithm, unlike the existing one, still works correctly under it.
\end{abstract}

\section{Introduction}

We study deterministic distributed consensus in synchronous dynamic
networks connected by unreliable, \emph{unidirectional} links. Assuming
unidirectional communication, in contrast to most existing research
\cite{KLO10:STOC,KOM11}, is not only of theoretical interest:
According to \cite{NKYG07}, 80\% of the links in a typical
wireless network are sometimes asymmetric. In fact, in wireless
settings
with low node density, various interferers and obstacles that severely inhibit 
communication, as in disaster relief applications \cite{LHSP11}, for example,
bidirectional links may simply not be achievable. Moreover, implementing 
low-level bidirectional communication between every pair of nodes is 
costly in terms of energy consumption, delay time and hardware 
resources. It may hence be an overkill for applications that just need
some piece of information available at one node to reach some other node,
as this is also achievable via directed multi-hop paths.
Obviously, in such settings, algorithmic solutions that do not 
assume bidirectional single-hop communication in the first place
provide significant advantages.

In this paper, we model directed dynamic networks
as synchronous distributed systems made up of $n$ 
processes, where processes have no knowledge of $n$.
In every round, the processes attempt a full message exchange
and compute a new local state based on the messages successfully received
in the message exchange. The actual communication in round $r=1,2,\dots$ 
is modeled as a sequence of directed communication graphs $\G^1, \G^2, \dots$,
which are considered under the control of a omniscient \emph{message 
adversary} \cite{AG13,RS13:PODC}: The \MA{} determines which messages 
are delivered and which get lost in each round.

In contrast to \cite{AG13}, where message adversaries are oblivious in
the sense that they can choose the round graphs arbitrarily from a \emph{fixed} 
set of candidates only, this paper, inspired by the research in
\cite{BRS12:sirocco,SWSBR15:NETYS},
considers message adversaries that may pick the graphs generated in some round 
depending on the particular round number. Obviously, this allows to model
\emph{stabilizing behavior}, which is not only of theoretical interest 
but also relevant from a practical point of view:
Starting-up a real dynamic distributed system
is likely a quite chaotic process, as nodes boot at different times 
and execute various initialization procedures. One can expect, though, that
the system will operate in a better orchestrated way after some unpredictable
startup time. A similar effect can be expected after a period of excessive transient 
faults, as caused by the abundant ionizing particles emitted during heavy solar flares 
\cite{Bau05,DR98}, for example. In this paper, we hence focus on stabilizing
message adversaries, which allow finite initial periods where arbitrary graphs may be generated.

The distributed computing problem considered in this paper is consensus.
A consensus algorithm ensures that all processes in the system 
eventually agree on a common decision value, which is computed 
(deterministically) from local inputs. It is an important primitive 
for any distributed application where data consistency is crucial.
Unlike in dynamic networks with unreliable \emph{bidirectional} links, where solving 
consensus is relatively easy \cite{KOM11}, solving consensus 
under message adversaries that generate unreliable directed
links is inherently difficult: For example, it is 
impossible to solve synchronous deterministic consensus
with two processes connected by a pair of lossy directional
links \cite{SW89}, even when it is guaranteed that only one link
can fail in every round \cite{SWK09}.
Therefore, in order to solve consensus, the power of the adversary must be
restricted somehow.
Exploring the solvability/impossibility-border
for consensus in directed dynamic networks is hence an interesting and
challenging topic.

\subsection*{Contributions} 

(1) We present two variants of
a ``natural'' stabilizing message adversary, which takes into consideration the
eventually stabilizing behavior that can reasonably be expected from real 
dynamic networks. During some finite initial period, the communication graphs
can be (almost) arbitrary: In particular, they may contain any
number of \emph{root
components}\footnote{Note that root components have already been
used in the asynchronous consensus algorithm for a minority of initially 
dead processes introduced by Fischer, Lynch and Paterson in \cite{FLP85}.}
(strongly connected
components that have no incoming edges from outside of the component),
which may even consist of the same set of nodes (with possibly varying
interconnect topology) for up to $D$ consecutive rounds.
$1\leq D < n$ is a system parameter, known to the processes, which
ensures that information from all members of a single root component
that remains the same for at least $D$ rounds reaches all $n$ processes
in the system. 
The ``chaotic'' initial period ends, at some unknown stabilization round $\rSR$, when,
for the first time, a single root component $R$ occurs that consists of the same
set of processes for more than $D$ consecutive rounds. 
%US: Das ist ein wenig misleading: Einen Broadcaster haben wir auch schon nach D round ...
%MS: da in der chaotic period auch partitioning erlaubt ist eigentlich nicht..
%This root component $R$ guarantees that eventually at least one process
%reaches all other processes in the system, i.e., guarantees a broadcaster.
%Naturally this is a necessary property to solve consensus.

The simple \emph{eventually stable forever after} variant of our
message adversary, $\EStable$, guarantees that $R$ remains a root component in all 
rounds after $\rSR$.
$\EStable$ is quite restricted in its behavior after stabilization, but 
is easy to analyze and facilitates an easy comparison of the performance 
(in particular, of the termination times) of different consensus algorithms.
The rigid properties of $\EStable$ are relaxed considerably in the
case of our message adversary $\AltEStable$, which just requires that $R$ 
re-appears, as a single root component, in at least $D$ (arbitrary, i.e., 
non-consecutive) rounds in the execution suffix after $\rSR+D$.

\medskip
\noindent
(2) We prove that no consensus algorithm can terminate under $\EStable$
(and hence under $\AltEStable$) before $\rSR+2D$. 
Note that the fastest known algorithm to date was presented in \cite{SWSBR15:NETYS} and
also works under $\EStable$. It has a termination 
time of $\rSR+4D$ and is hence sub-optimal here.

\medskip
\noindent
(3) We provide a simple consensus algorithm, which matches the termination
time lower bound of $2D+1$ under $\EStable$ and works correctly also under
$\AltEStable$. Note that the algorithm from \cite{SWSBR15:NETYS} fails 
under $\AltEStable$, even though its code is considerably more complex.

\subsection*{Previous results}

In \cite{BRS12:sirocco}, Biely et.al.\ showed that
consensus is solvable under a message adversary that generates graphs
containing a single root component only, which eventually
consists of the same processes for at least $4D$ consecutive rounds;
the term \emph{$4D$-vertex-stable root component} has been coined
to reflect this fact.
Note that vertex-stable root components neither imply a static network nor a 
stable subgraph over multiple rounds.
It has also been shown in \cite{BRS12:sirocco} that consensus is impossible
if the adversary is not forced to generate a root component that is vertex-stable
for at least $D$ rounds.

In \cite{SWSBR15:NETYS}, we showed 
 that consensus can be solved under a message adversary
that may generate multiple vertex-stable roots,
albeit with a worse worst case termination time and a far more
complex algorithm. More specifically, the \MA{} proposed in this paper
guarantees root components that
(i) are eventually stable
for at least $4D$ rounds concurrently, and (ii) ensures some distinct information
flow between successive vertex-stable root components (``majority influence''). The proposed
algorithm is gracefully degrading, in the sense that it solves $k$-set
agreement for the worst-case optimal choice of $k$, when consensus ($k=1$)
cannot be solved in the given run.
Recall that in $k$-set agreement, the consensus agreement condition is relaxed such that
up to $k$ different decision values are permitted.

\subsection*{Other related work}

Dynamic networks have been studied intensively in distributed computing
(see the overview by Kuhn and Oshman \cite{KO11:SIGACT} and the references therein). 
Besides work on peer-to-peer networks like \cite{KSW10},
where the dynamicity of nodes (churn) is the primary concern, different
approaches for modeling dynamic connectivity have been proposed, both
in the networking context and in the context of classic distributed
computing. $T$-interval-connectivity in synchronous distributed computations 
has been introduced in \cite{KLO10:STOC}. 

Agreement problems in dynamic networks with undirected 
communication graphs have been studied in the work by 
Kuhn et al.~\cite{KOM11}; it focuses on the $\Delta$-coordinated 
consensus problem, which extends consensus by requiring all processes to decide within
$\Delta$ rounds of the first decision. 
Agreement in directed graphs has been considered in 
\cite{SWK09,BRS12:sirocco,AG13,RS13:PODC,CG13,SWSBR14:PODC,SWSBR15:NETYS}. 
Whereas 
\cite{SWK09,CG13} considerably restrict the dynamicity of the 
communication graphs, e.g., by not allowing stabilizing behavior,
which effectively causes them to belong to quite strong classes of
network assumptions in the classification of Casteigts et al.~\cite{CFQS12:TVG},
the algorithms of \cite{BRS12:sirocco,SWSBR14:PODC,SWSBR15:NETYS} allow to
solve consensus under very weak network assumptions: \cite{BRS12:sirocco}
only admits single-rooted graphs, whereas \cite{SWSBR15:NETYS} provides
a consensus algorithm that gracefully degrades to $k$-set agreement in
unfavorable runs under a fairly strong stabilizing message adversary.
Afek and Gafni \cite{AG13} introduced (oblivious) message adversaries 
for specifying network assumptions in this context, and used them 
for relating problems solvable in wait-free read-write shared memory systems
to those solvable in message-passing systems. 
Raynal and Stainer \cite{RS13:PODC} used message adversaries for
exploring the relationship between round-based models and failure 
detectors. 

\section{Model} \label{sec:model}

%\subsubsection*{Principles of communication and computation}

We model a synchronous message passing system as a set
$\Pi$ of $|\Pi| = n > 1$ deterministic state machines, called
\emph{processes}.
Processes do not necessarily know $n$ but have unique identifiers 
that we pick, w.l.o.g., from the
set $\{ 1, \dots, n \}$.
In our analysis,
we use a process and its identifier interchangeably when
there is no ambiguity.
Processes operate in lock-step rounds, where each round consists of a phase
of full message exchange, followed by an instantaneous local \emph{computing step}.
Following \cite{SWSBR15:NETYS,BRS12:sirocco}, the actual communication in 
round $r \geq 1$ is according to a digraph\footnote{
Usually, we sloppily write $p \in \Gr$, resp.\ 
$(p \ra q) \in \Gr$ instead of $p \in V$ resp.\ $(p \ra q) \in E^r$.
}
$\Gr = (V, E^r)$ controlled by an omniscient \emph{message adversary}:
Each vertex in $V$ corresponds to exactly one process of $\Pi$, and an edge
from $p$ to $q$, denoted
$(p \ra q)$, is present in $E^r$ iff the adversary permits the delivery
of the message sent from $p$ to $q$ in round $r$. % (see \cref{fig:smp} for an example).
We assume that $\Gr$ contains self-loops $(p \ra p)$ for all $p \in V$,
i.e., processes always
receive their own message in every round.
Rounds are communication-closed, i.e., messages sent in some round $r$ and
delivered in a later round $r' > r$ are dropped.
\begin{comment}
\begin{figure}
 \begin{subfigure}[b]{0.30\textwidth}
  \centering
  \begin{tikzpicture}
   \draw (0,0) node (p1) {$p_1$};
   \draw (0.6,1) node (p2) {$p_2$};
   \draw (1.2,0) node (p3) {$p_3$};
   \draw (1.8,1) node (p4) {$p_4$};
   \draw (2.4,0) node (p5) {$p_5$};
   
   \draw[dot] (p1) to (p2); 
   \draw[dot] (p2) to (p3); 
   \draw[dot] (p3) to (p4); 
   \draw[dot] (p3) to (p1); 
   \draw[dot] (p4) to (p5); 
  \end{tikzpicture}
  \caption*{$\G^1$ for round $1$}
 \end{subfigure}
 \begin{subfigure}[b]{0.30\textwidth}
  \centering
  \begin{tikzpicture}
   \draw (0,0) node (p1) {$p_1$};
   \draw (0.6,1) node (p2) {$p_2$};
   \draw (1.2,0) node (p3) {$p_3$};
   \draw (1.8,1) node (p4) {$p_4$};
   \draw (2.4,0) node (p5) {$p_5$};
   
   \draw[dot] (p2) to (p3); 
   \draw[dot] (p3) to (p4); 
   \draw[dot] (p3) to (p1); 
  \end{tikzpicture}
  \caption*{$\G^2$ for round $2$}
 \end{subfigure}
 \begin{subfigure}[b]{0.30\textwidth}
  \centering
  \begin{tikzpicture}
   \draw (0,0) node (p1) {$p_1$};
   \draw (0.6,1) node (p2) {$p_2$};
   \draw (1.2,0) node (p3) {$p_3$};
   \draw (1.8,1) node (p4) {$p_4$};
   \draw (2.4,0) node (p5) {$p_5$};
   
   \draw[dot] (p3) to (p2); 
   \draw[dot] (p3) to (p4); 
   \draw[dot] (p3) to (p1); 
   \draw[dot] (p3) to (p5); 
  \end{tikzpicture}
  \caption*{$\G^3$ for round $3$}
 \end{subfigure}
 \caption{Communication graphs $\G^1,\G^2,\G^3$ for 3 rounds of synchronous message passing.} \label{fig:smp}
\end{figure}
\end{comment}

The messages sent and the state transitions performed by the processes 
in a round are guided by a deterministic message-sending and 
state-transition function, respectively, which are specified implicitly
by algorithms in pseudo-code: 
 The \emph{local state} of a process comprises 
all its local variables; the \emph{message-sending function} determines
the message to be broadcast in a round, and the \emph{state-transition function}
determines the local state reached at the end of the round, depending
on the previous state and the set of messages received in the round.
Most of the time, we will assume that the algorithms are
\emph{full-information}, i.e.,
processes keep track of received messages and forward their entire states to all processes they can reach in every
round.
%\footnote{We comment on efficient algorithms in \cref{sec:refinements}.}

In our analysis, $\state{p}{r}$ denotes
the \emph{local state} of process $p$ at the end of round $r\geq 1$, after its computing
step; $\state{p}{0}$ is the initial state at the beginning of round 1.
The value of a particular variable $\mathit{var}$ in $\state{p}{r}$ 
is denoted by $\mathit{var}_p^r$.\footnote{Note that, 
throughout our paper, superscripts usually denote round numbers, with the implicit
assumption that they refer to the end of a round (after the computing step), whereas subscripts 
typically identify processes.} The vector of states of all the processes at the end 
of round $r$ is called round $r$ \emph{configuration} $C^r$; $C^0$ denotes the initial configuration.
An \emph{execution}, or \emph{run}, is an alternating sequence of configurations and communication 
graphs. As our algorithms are deterministic, it is uniquely determined by a given initial configuration $C^0$ together with 
an infinite sequence\footnote{As usual, we denote by $\Seqr{a}{b}$ the sequence
$(\G^a, \ldots, \G^b)$ of communication graphs.}
of communication graphs $\SeqInf$, which is controlled by a \emph{\MA{}}.
%When talking about the round $r$ state of process $p$ in a specific execution $\varepsilon$,
%we use $\state{p^\varepsilon}{r}$. 
More generally, any execution segment, starting 
from configuration $C^r$, is uniquely specified by a tuple like
$\Exec{ C^r, \Seq{i}{r+1}{a}, \Seq{j}{a+1}{b}, \ldots}$.
An execution is called \emph{admissible}, if it is in accordance with the message-sending
and state-transition functions of the processes and the definition of the \MA{}.

As in \cite{SWSBR15:NETYS}, we will restrict the power of a \MA{} in terms of the properties of the
sequences of communication graphs it may legitimately generate.
Consequently, an adversary $A$ that has a set of properties $P_A$ can formally be 
specified via the set of its \emph{feasible} infinite
communication graph sequences $A := \{ \SeqInf \, | \, \SeqInf \: \mbox{satisfies} \: P_A \}$.
We say that an adversary $A$ is weaker than an adversary $B$, resp.\ that $B$
is stronger than $A$, if all feasible sequences of $A$ are also in $B$ but not
vice-versa, i.e., $A \subset B$.
If $A$ contains sequences not in $B$ and $B$ contains sequences not in $A$,
$A$ and $B$ are incomparable. An example for two incomparable adversaries is the adversary that
allows only chains for each $\Gr$ and the adversary that allows only
circles for each $\Gr$.

\begin{comment}
\begin{figure}
	\centering
	$\mathtt{ONLY\text{-}CHAINS} \subset \mathtt{ONLY\text{-}TREES}$\\
	$\not\subseteq, \not\supseteq$ \\
	$\mathtt{ONLY\text{-}CIRCLES} \subset \mathtt{ONLY\text{-}SCC}$
\caption{Example for a relation between \MAs{}. \todo{Draft}}
\label{fig:adv}
\end{figure}
\end{comment}

We say that a problem is \emph{impossible} under some \MA{} if
there is no deteministic algorithm that solves the problem for every feasible
communication graph sequence. For example, every problem that requires
at least some communication among the processes is impossible under the
unrestricted \MA{}, which may generate
all possible graph sequences: The sequence $\SeqInf$ where no $\Gr$ 
contains even a single edge is also feasible here.

We are interested in solving the \emph{consensus problem}, where each process $p$
has an initial value $x_p$ and a write-once decision value $y_p$ in its local state.
Formally, the following conditions must be met in every execution of a correct consensus
algorithm in our setting 
for $p, q \in \Pi$:
\begin{compactitem}
   \item[(Agreement)] If $p$ assigns value $v_p$ to $y_p$ and
                      $q$ assigns $v_q$ to $y_q$, then $v_p = v_q$.
   \item[(Termination)] Eventually, every $p$ assigns a value to $y_p$.
   \item[(Validity)] If $p$ assigns a value $v$ to $y_p$, then there is some $q$ such that 
$x_q = v$.
\end{compactitem}

%\todo{do we need bivalence?}
%We call the special case of consensus where the set of initial values is
%restricted to $\{ 0, 1 \}$ binary consensus.
%Let $\A$ be a binary consensus algorithm under \MA{} and
%$v \in \{ 0, 1 \}$.
%Analogous to \cite{FLP85}, we call a round $r$ configuration $C_r$ of $\A$
%$v$-valent if no graph sequence $\Seq{i}{r+1}{s}$ exists such that some
%process $p$ assigns $1-v$ to $y_p$ in $\Exec{C_r, \Seq{i}{r+1}{s}}$.
%A configuration that is not $v$-valent for some $v \in \{0, 1 \}$ is called
%bivalent.

\subsection*{Dynamic graph concepts}

As in \cite{SWSBR15:NETYS,BRS12:sirocco}, the message adversaries considered
in this paper will focus on \emph{root components} in the communication graphs, which
are strongly connected components that have no incoming edges. Their importance
has already been recognized in the celebrated paper \cite{FLP85}
by Fischer, Lynch and Paterson, which also introduces an algorithm for asynchronous consensus
with a minority of initially dead processes. It essentially
identifies the (unique) root component in the initial communication graph formed by
the processes waiting for first $n/2$ messages to arrive.

\begin{definition}[Root Component]
  A non-empty set of nodes $R \subseteq V$ is called a round $r$ root
  component of $\Gr$, if it is the set of vertices of a strongly connected 
  component $\cal R$ of $\Gr$ and
  $\forall p \in \Gr, q \in R : (p \ra q) \in \Gr \Rightarrow p \in R$.
  We denote by $\roots(\Gr)$ the set of all root components of $\Gr$, resp.\
  the single root component of $\Gr$, and by $|R|$ the number of nodes in $R$.
\end{definition}

By contracting the strongly connected components of $\Gr$, it is easy
to see that every graph has at least one root component (just called ``roots'' for brevity).
Furthermore, if $\Gr$ contains a single root only, contraction
leads to a tree, so $\Gr$ must be weakly connected in this case.

\begin{corollary}
  For any directed graph $\Gr$,
  $\abs{\roots(\Gr)} \geq 1$, and if $\abs{\roots(\Gr)}=1$, then $\Gr$ is weakly connected.
  \label{cor:root-properties}
\end{corollary}

We call a
set of nodes $R$ that forms a root component in every communication graph of
a sequence $\SeqrI$ a \emph{common root}
of this sequence. Note carefully
that the interconnect topology of the nodes in $R$, i.e., the root component 
$\cal R$ taken as a subgraph of $\Gr$, as well as the outgoing edges to the remaining 
nodes $\Pi\setminus R$ in $\Gr$, may be different in every round $r$ in the sequence.
The index set $I$ of rounds in $\SeqrI$ is usually an interval $I=[a,b]$ of 
$|I|=b-a+1$ consecutive 
rounds\footnote{In \cite{BRS12:sirocco,SWSBR15:NETYS}, the term
\emph{$I$-vertex-stable root component} ($I$-VSRC, or alternatively $d$-VSRC)
has been coined for $R$ being a common root in $\SeqrI$ with $I=[a,a+d-1]$.
We prefer the more general term common root of a sequence in this paper, since
it aligns better with the focus of our analysis on (possibly arbitrary) sequences of
communication graphs.} (we will call $\SeqrI$ a \emph{consecutive} graph sequence in this
case), but can also be an arbitrary index set that is ordered 
according to increasing round numbers. 
If a consecutive graph sequence is maximal wrt.\ $R$ being its common root, we call 
$R$ a maximal common root.

\begin{definition}[Common root]\label{def:common-root}
  We say that a sequence $\SeqrI$ has a common root $R$, iff there
  exists a root $R$ (with possibly different interconnect topology) 
  such that $R \in \roots(\Gr)$ for all $r \in I$. If $I=[a,b]$ with 
$|I|=b-a+1$ is an interval of consecutive rounds $a,a+1,\dots,b$,
$\SeqrI$ is called a consecutive graph sequence. We call $R$ a maximal 
common root of a consecutive graph sequence $\Seqr{a}{b}$, iff $R$ is a common root
of $\Seqr{a}{b}$ but neither of $\Seqr{a-1}{b}$ nor $\Seqr{a}{b+1}$.
\end{definition}

Finally, a graph sequence that has a unique common root is called a
single-rooted sequence.

\begin{definition}[Single-rooted sequence]
We call a sequence $\SeqrI$ single-rooted, or $R$-single-rooted,
if there exists a unique root component $R$ s.t.\
$\forall i, j \in I : \roots(\G^i) = \roots(\G^j) =  \set{R}$.
We call $R$ a maximal single root of a consecutive graph sequence 
$\SeqrI$ with $I=[a,b]$, iff $R$ is a single root
of $\Seqr{a}{b}$ but neither of $\Seqr{a-1}{b}$ nor $\Seqr{a}{b+1}$.
\label{def:single-root}
\end{definition}

We now introduce a notion of \emph{causal past}, which is
closely related to the classic ``happens-before'' relation \cite{Lam78},
albeit presented in a way that is compatible with the process-time 
graphs used e.g.\ in \cite{KOM11}.
Given some round $b$, $p$'s \emph{causal past} $\CP{p}{a}{b}$
down to round $a$ are exactly 
those processes the state of which at the end of round $a$
has affected the state of $p$ at the end of round $b$.

\begin{definition}[Causal past]\label{def:causal-influence}
For a given infinite sequence $\sigma$ of communication graphs,
%let the causal future $\CF_p(a,b)$
%of process $p$ from (the end of) round $a$ up to (the end of) round $b$ 
%be inductively defined as:
%$\CF_p(a, a) := \{ p \}$ and for $a \leq \ell < b$,
%$\CF_p(a, \ell+1) := \CF_p(a, \ell) \cup \{q \in \Pi |
%\exists q' \in \CF_p(a, \ell) :
%(q' \ra q) \in \G^{\ell+1}
% \}$.
%Analogously,
we define the causal past $\CP{p}{a}{b}$
of process $p$ from (the end of) round $b$ down to (the end of) round $a$ as
$\CP{p}{b}{b} := \set{p}$ and for $a < \ell \leq b$,
$\CP{p}{\ell-1}{b} := \CP{p}{\ell}{b} \cup \{q \in \Pi \mid
\exists q' \in \CP{p}{\ell}{b} :
(q' \ra q) \in \G^{\ell}
 \}$
%
%Given $\sigma$, we say that process \emph{$p$ (at the end of round $a$) causally influences $q$ (at the end of round $b$)}, denoted by
%$\state{p}{a} \leadsto_{\sigma} \state{q}{b}$,
%or simply $\infl{p}{a}{q}{b}$ when
%$\sigma$ is understood,
%if $q \in \CF_p(a,b)$, or, equivalently, if $p \in \CP{q}{a}{b}$.
\end{definition}

Note carefully an important consequence of \cref{def:causal-influence}: By definition,
$\infl{q}{a}{p}{b}$ implies that the state of $q$ at the end of round $a$ is in the causal
past of $p$ by the end of round $b$.
Since the latter is a direct result of the communication graphs up to round $b$,
however, this implies that $p$ must have
got the information about the round~$a$ state of $q$ already \emph{before} it performs its
round $b$ computing step, e.g., in a round~$b$ message.
Thus, $p$ can use that information already in its round $b$ computation.

From the monotonic growth of $\CP{p}{a}{b}$ (recall the self-loops in every $\Gr$), 
we can deduce the following corollary:

\begin{corollary} \label{cor:transitive-influence}
$\infl{p}{a}{q}{b}$ implies
$\infl{p}{a}{q}{b'}$
for all $b' \geq b$.
Analogously,
$\infl{p}{a}{q}{b}$ implies that
$\infl{p}{a'}{q}{b}$ for all $a' \leq a$.

\end{corollary}

As it will turn out in the next section, the ``multi-hop delay'' 
of a message sent by some process to reach some other process(es),
i.e., the speed of information propagation over multiple rounds,
will be important for solving consensus.
This is particularly true in the case of a single-rooted graph
sequence, where the following lemma guarantees an upper bound
of $n-1$ rounds:

\begin{lemma}
Let $\sigma$ be a graph sequence containing a sequence
$S = \left(\G^{r_1}, \ldots, \G^{r_{n-1}} \right)$
of $n-1$ not necessarily consecutive
$R$-single-rooted communication graphs. 
Then, for all $p \in \Pi : \setinfl{R}{r_1-1}{p}{r_{n-1}}$.
\label{cor:end-to-end}
\end{lemma}

  \begin{proof}
  Pick an arbitrary process $p \in \Pi$, $q \in R$.
  We show by induction that, for $\ell \in [1, n-1]$,
  $|\CP{p}{r_{n-\ell}}{ r_{n-1}}| \geq \ell$ or 
  $q \in \CP{p}{r_{n-\ell}}{ r_{n-1}}$.
  For $\ell = 1$, this follows directly from \cref{def:causal-influence}.
  For the induction step, we assume that the claim holds for
  $\ell \in [1, n-1)$ and show that it holds for $\ell+1$ as well.
  If the claim holds because $q \in \CP{p}{r_{n-\ell}}{ r_{n-1}}$,
  by \cref{cor:transitive-influence},
  we have $q \in \CP{p}{r_{n- \ell -1}}{ r_{n-1}}$.
  Thus, assume that $q \notin \CP{p}{r_{n-\ell}}{ r_{n-1}}$ and
  $|\CP{p}{r_{n-\ell}}{ r_{n-1}}| \geq \ell$.
  If it holds that $|\CP{p}{r_{n-\ell}}{ r_{n-1}}| > \ell$, we get
  $|\CP{p}{r_{n-\ell-1}}{ r_{n-1}}| \geq \ell+1$ immediately,
  so assume that $|\CP{p}{r_{n-\ell}}{ r_{n-1}}| = \ell$.
  Since $\G^{r_{n-\ell}}$ is $R$-single-rooted, there is a path from
  $q$ to $p$ in $\G^{r_{n-\ell}}$, according to \cref{cor:root-properties}.
  Because $q \notin \CP{p}{r_{n-\ell}}{ r_{n-1}}$, there is some process
  $q'$ on the path from $q$ to $p$ s.t.\
  $q' \notin \CP{p}{r_{n-\ell}}{ r_{n-1}}$ but
  $(q' \ra p') \in \G^{r_{n-\ell}}$ for some
  $p' \in \CP{p}{r_{n-\ell}}{ r_{n-1}}$.
  By \cref{def:causal-influence},
  $\CP{p}{r_{n-\ell-1}}{r_{n-1}} \supseteq \CP{p}{r_{n-\ell}}{r_{n-1}} \cup \set{q'}$.
  By the induction hypothesis, therefore $|\CP{p}{r_{n-\ell-1}}{ r_{n-1}}| \geq \ell + 1$.
\end{proof}

In order to specify message adversaries that guarantee faster information
propagation than guaranteed by \cref{cor:end-to-end}, we introduce a system
parameter called \emph{dynamic (network) diameter} $1\leq D\leq n-1$. 
Intuitively, it ensures
that the information from all nodes in $R$ has reached all nodes in the 
network if $D$ $R$-single-rooted graphs have occurred in a graph sequence.

\begin{definition}[Dynamic diameter D]\label{def:dynamic-diameter}
A message adversary $\MAD$ guarantees a dynamic (network) diameter $D$,
if for every graph sequence $\sigma \in \MAD$ that contains a subsequence
$S = \left(\G^{r_1}, \ldots, \G^{r_D} \right)$
of $D$ not necessarily consecutive $R$-single-rooted communication graphs,
it holds that $R \subseteq \CP{p}{r_1-1}{r_{D}}$ for every $p\in\Pi$.
\end{definition}

It was shown in \cite[Theorem 3]{BRS12:sirocco}
that processes need to know some estimate of $D$ for solving consensus:
Without this knowledge, it is impossible to locally verify a necessary
condition for solving consensus, namely, the ability of some process to
disseminate its initial value system-wide. Note carefully, though, that
knowledge of $D$ does not permit the processes to determine $n$ in general.

\iftoggle{TR}{
\Cref{def:dynamic-diameter} may lead to the conjecture that a maximum hop distance 
of $D$ between $q\in R$ and $p \in \Pi$ in every $\G^{r_1}, \dots, \G^{r_D}$
guarantees a dynamic diameter of $D$.
This is not the case, however: Consider, for example, the three-round sequence
$\Seqr{1}{3}$ of communication graphs for processes $p_1, \ldots, p_5$ shown
in \cref{fig:bc-delay}. Herein, $\G^1$ is a directed tree of height~3, 
with single root node $p_1$ and a single node in the second level.
In the following rounds, this second level node switches places with a new node $\neq p_5$
from the third level.
In this scenario, $\ninfl{p_1}{0}{p_5}{3}$,
even though the length of the path from
$p_1$ to any other process is $\leq 2$ in every $\Gr$.

\begin{figure}
  \centering
  \begin{tikzpicture}
    \draw(0,0)     node[proc] (p1) {$p_1$};
    \draw(0,-1)    node[proc] (p2) {$\mathbf p_2$};
    \draw(-1,-2)   node[proc] (p3) {$p_3$};
    \draw(0,-2)    node[proc] (p4) {$p_4$};
    \draw(1,-2)    node[proc] (p5) {$\mathbf p_5$};
    
    \draw(0,-2.7)    node[proc] (l)   {\scriptsize Round 1};

    \draw[dot] (p1) to (p2);
    \draw[dot] (p2) to (p3);
    \draw[dot] (p2) to (p4);
    \draw[dot] (p2) to (p5);

    \draw(3,0)   node[proc] (p1) {$p_1$};
    \draw(3,-1)  node[proc] (p2) {$\mathbf p_3$};
    \draw(3,-2)  node[proc] (p4) {$p_4$};
    \draw(2,-2)  node[proc] (p3) {$\mathbf p_2$};
    \draw(4,-2)  node[proc] (p5) {$\mathbf p_5$};
    
    \draw(3,-2.7)    node[proc] (l)   {\scriptsize Round 2};

    \draw[dot] (p1) to (p2);
    \draw[dot] (p2) to (p3);
    \draw[dot] (p2) to (p4);
    \draw[dot] (p2) to (p5);

    \draw(6,0)   node[proc] (p1) {$p_1$};
    \draw(6,-1)  node[proc] (p2) {$\mathbf p_4$};
    \draw(6,-2)  node[proc] (p4) {$\mathbf p_3$};
    \draw(5,-2)  node[proc] (p3) {$\mathbf p_2$};
    \draw(7,-2)  node[proc] (p5) {$\mathbf p_5$};
    
    \draw(6,-2.7)    node[proc] (l)   {\scriptsize Round 3};

    \draw[dot] (p1) to (p2);
    \draw[dot] (p2) to (p3);
    \draw[dot] (p2) to (p4);
    \draw[dot] (p2) to (p5);

  \end{tikzpicture}

 \caption{Example of a communication graph sequence with dynamic diameter $D =4$, despite a small
 hop distance (diameter = 2) in every single graph. Bold nodes represent processes in the causal past $\CP{p_5}{0}{3}$.}
 \label{fig:bc-delay}
\end{figure}
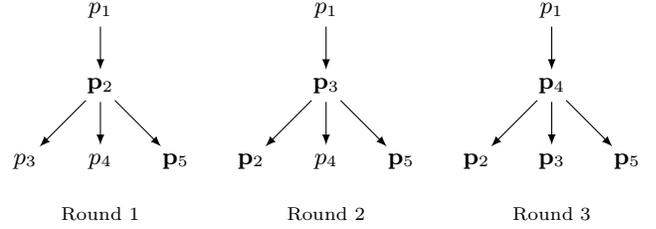
}{}

\section{A simple stabilizing message adversary} \label{sec:adversary}

Recall that the purpose of our stabilizing message adversary is to allow
an unbounded (but finite) initial period of ``chaotic'' behavior, where
the communication graphs can be arbitrary: Unlike in \cite{BRS12:sirocco},
any $\Gr$ may be arbitrarily sparse and could contain several root components here. Clearly,
one cannot hope to solve consensus during this initial period in general. Eventually,
however, the adversary must start to generate suitably restricted 
communication graphs, which should allow the design of algorithms that 
solve consensus. Naturally, there are many conceivable restrictions and,
hence, many different message adversaries that could be considered here.
We will develop two instances in this paper, and also relate those to
the \MA{} introduced in \cite{SWSBR15:NETYS}.

%We will define our message adversary via the properties of the round graphs 
%$\Gr$ of selected sub-sequences $\SeqrI \subseteq \SeqInf$, as
%this allows a natural and fine-grained specification.

The simple message adversary $\EStable$ defined in this section uses a 
straightforward means for closing the initial period, which is 
well-known from eventual-type models in distributed computing:
In partially synchronous systems \cite{DLS88}, for example, one
assumes that speed and communication delay bounds hold forever 
from some unknown stabilization time on. Analogously, we
assume that there is some unknown round $\rGST$, from which on
the adversary must behave ``nicely'' forever. Albeit the resulting
message adversary is restricted in its behavior, it
provides easy comparability of the
performance (in particular, of the termination times) of different 
consensus algorithms. Moreover, in \cref{sec:refinements}, we will show how
to generalize $\EStable$ to a considerably stronger message
adversary $\AltEStable$, which does not require such a restrictive ``forever
after'' property.

\def\SeqrIGST{\Seqr{\rGST}{\infty}}
In order to define what ``behaving nicely'' actually 
means in the case of $\EStable$, we start from a necessary condition for solving 
consensus in $\SeqrIGST$: The arguably most obvious requirement 
here is information propagation from a non-empty set of processes 
to all processes in the system. According to \cref{cor:end-to-end},
this can be guaranteed when there is a sufficiently long
sub-sequence of communication graphs in $\SeqrIGST$ with a 
single common root. Natural 
candidate choices for feasible graphs would hence be the very same 
single-rooted graph $\G$ in
all rounds $r\geq \rGST$, or the assumption that all $\Gr$ are
strongly or even completely connected (and hence also single-rooted).
While simple, these choices would impose severe and unnecessary restrictions on our \MA,
however, which are avoided by the following more general definition (that includes
these choices as special instances, and hence results in a stronger \MA):

\begin{definition} \label{def:liveness}
We say that $\SeqInf$ has a (unique) \emph{\FAES-common root $R$} (``forever after,
eventually single'') starting at round $\rGST \geq 1$,
iff $R$ is (i) a maximal common root of $\Seqr{\rGST}{\infty}$ and 
(ii) a maximal single root of $\Seqr{\rSR}{\infty}$, for some round $\rSR \geq \rGST$.

$\Liveness$ contains those communication graph sequences $\SeqInf$
that have a \FAES-common root $R$.
\end{definition}

Note that the eventual single-rootedness of $\SeqrIGST$ implied by 
$\Liveness$ allows the respective round graphs $\Gr$ to be very sparse:
For instance, each $\Gr$ of $\SeqrIGST$ consisting of a chain
with the same head but varying body would satisfy the requirement for 
single-rootedness. 

Whereas the properties guaranteed by $\Liveness$ will suffice to ensure liveness 
of the consensus algorithm presented in \cref{sec:algorithm}, i.e., termination,
it is not sufficient for also ensuring safety, i.e., agreement.
\iftoggle{TR}{
Consider for instance the top run (execution $\varepsilon_1$) from
\cref{fig:stab-is-not-enough},
where $p$ is connected to $q$ in a chain
forever, which is feasible for $\Liveness$.}
{Consider for instance the execution $\varepsilon_1$,
where $p$ is connected to $q$ in a chain
forever, which is feasible for $\Liveness$.}
In any correct solution algorithm, the head $p$ of this chain must
eventually decide in some round $\tau$ on its initial value $x_{p}$.
\iftoggle{TR}{
Now consider the execution $\varepsilon_2$, depicted in the
bottom of \cref{fig:stab-is-not-enough}, where $p$
is disconnected until $\tau$
and $x_{p} \neq x_{q}$.}
{Now consider the execution $\varepsilon_2$, where $p$
is disconnected until $\tau$
and $x_{p} \neq x_{q}$.}
Since $\varepsilon_2$ is indistinguishable for $p$ 
from $\varepsilon_1$ until $\tau$, process $p$ will decide $x_{p}$ at time 
$\tau$. However, in $\varepsilon_2$, a chain forms with head $q \neq p$ 
 forever after $\tau$.
Since $q$ is only aware of its own input value $x_q$, it can never 
make a safe decision in this execution.

\iftoggle{TR}{ \begin{figure}
  \centering
  \begin{tikzpicture}[]
    \draw (0,0)    node[proc] (p) {$p$}; 
    \draw (1,0)    node[proc] (q) {$q$}; 
    \draw (0,-1.5) node[proc] (p') {$p$}; 
    \draw (1,-1.5) node[proc] (q') {$q$}; 

    \draw[dot] (p) to (q);

    \draw (0.5, -2.2) node (l) {\scriptsize Rounds $1$ to $\tau$};

    \draw (3,0)    node[proc] (p1) {$p$}; 
    \draw (4,0)    node[proc] (q1) {$q$}; 
    \draw (3,-1.5) node[proc] (p1') {$p$}; 
    \draw (4,-1.5) node[proc] (q1') {$q$}; 

    \draw[dot] (p1) to (q1);
    \draw[dot] (q1') to (p1');

    \draw (3.5, -2.2) node (l) {\scriptsize Rounds $\tau+1$ to $\infty$};
  \end{tikzpicture}
 \caption{Two executions $\varepsilon_1$ (top) and $\varepsilon_2$ (bottom), 
indistinguishable for $p$ until $\tau$.}
 \label{fig:stab-is-not-enough}
\end{figure}
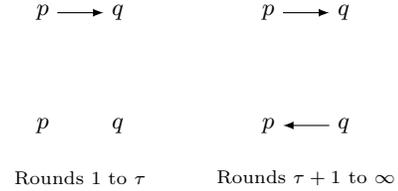

}{}

This is why $\EStable$ needs to combine $\Liveness$ with another message
adversary $\Safety(x)$ that enables our solution algorithm to also 
ensure safety. The above example illustrates the main problem that we face here:
If we allow root components to remain common for too many consecutive rounds
in the initial period (before $\rGST$), the members of such a root component (which does not need
to be single) cannot distinguish
this from the situation where they are belonging to the final \FAES-common root 
(after $\rGST$).
In \cite{BRS12:sirocco}, this problem was void since \emph{all} communication
graphs were assumed to be single-rooted. In the following \cref{def:safety}, 
we require that
every root $R$ that is common during a sequence of ``significant'' 
length $x+1$ is already the \FAES-common root $R$. Again, in \cref{sec:refinements}, 
we will present a significant relaxation of this quite restrictive (but convenient)
assumption.

\begin{definition}\label{def:safety}
$\Safety(x)$ contains those communication graph sequences $\sigma=\SeqInf$, where
every root $R$ that is common for $> x$ consecutive rounds in $\sigma$
is the \FAES-common root $R$ in $\sigma$.
\end{definition}

We are now ready to define our simple eventually stabilizing \MA{}
$\EStable$, which is the conjunction of the adversaries from \cref{def:liveness,def:safety},
augmented by the additional requirement to always
guarantee a dynamic network diameter $D$ according to \cref{def:dynamic-diameter}:

\begin{definition} \label{def:EStableMA}
The \MA{} $\EStable =$\\
$\Safety(D) + \Liveness$ contains those graph sequences of
$\Safety(D) \, \cap \, \Liveness$ that guarantee a dynamic diameter of $D$.
\end{definition}

For exemplary graph sequences of $\EStable$ with $D=2$, see
\cref{fig:EStable-impossibility-eps1,fig:EStable-impossibility-eps2}.
Note carefully that \cref{def:liveness} allows the coexistence of
the \FAES-common root $R$ with some other root component $R' \neq R$
in communication graphs that occur before
$R$ becomes the single root (in round $\rSR$).
However, according to \cref{def:safety},
$R'$ cannot be common root for more than $D$ consecutive rounds in
this case.

\begin{comment}
\todo{Maybe figures/examples of some executions admissible under $\EStable$?}
\begin{figure}
 \begin{subfigure}[b]{0.30\textwidth}
  \centering
  \begin{tikzpicture}
   \draw (0,0) node (p1) {$p_1$};
   \draw (0.6,1) node (p2) {$p_2$};
   \draw (1.2,0) node (p3) {$p_3$};
   \draw (1.8,1) node (p4) {$p_4$};
   \draw (2.4,0) node (p5) {$p_5$};
   
   \draw[dot] (p1) to (p2); 
   \draw[dot] (p2) to (p3); 
   \draw[dot] (p3) to (p1); 
   \draw[dot] (p4) to (p5); 
  \end{tikzpicture}
  \caption*{Rounds $1$ to $D-3$}
 \end{subfigure}
 \begin{subfigure}[b]{0.30\textwidth}
  \centering
  \begin{tikzpicture}
   \draw (0,0) node (p1) {$p_1$};
   \draw (0.6,1) node (p2) {$p_2$};
   \draw (1.2,0) node (p3) {$p_3$};
   \draw (1.8,1) node (p4) {$p_4$};
   \draw (2.4,0) node (p5) {$p_5$};
   
   \draw[dot] (p1) to (p2); 
   \draw[dot] (p2) to (p3); 
   \draw[dot] (p3) to (p1); 
   \draw[dot] (p3) to (p4); 
   \draw[dot] (p4) to (p5); 
  \end{tikzpicture}
  \caption*{$D-2$ to $2D+1$}
 \end{subfigure}
 \begin{subfigure}[b]{0.30\textwidth}
  \centering
  \begin{tikzpicture}
   \draw (0,0) node (p1) {$p_1$};
   \draw (0.6,1) node (p2) {$p_2$};
   \draw (1.2,0) node (p3) {$p_3$};
   \draw (1.8,1) node (p4) {$p_4$};
   \draw (2.4,0) node (p5) {$p_5$};
   
  \end{tikzpicture}
  \caption*{$2D+2$ to $\infty$}
 \end{subfigure}
 \caption*{Admissible run under $\EStable$ \todo{More (representative) executions?}}
 \label{fig:arun}
\end{figure}
\end{comment}

\iftoggle{TR}{
  \iftoggle{TR}{ In the remainder of this section, we will}{We
now briefly and} informally introduce the 
message adversary $\mathtt{VSRC}(n, 4D) + \mathtt{MAJINF}(k)$ introduced 
in \cite{SWSBR15:NETYS}.\footnote{In \cite{SWSBR15:NETYS},
a network diameter $H$ and a root diameter $D$
are distinguished; we set $H=D$ here to ensure compatibility with
our definitions.} The latter paper
introduced a consensus algorithm, which gracefully degrades to $k$-set
agreement\footnote{In $k$-set agreement, the consensus agreement condition 
is relaxed such that up to $k$ different decision values are permitted.} 
in less favorable runs. 
$\mathtt{VSRC}(n, 4D)$ consists of all graph sequences $\sigma$, where up to $n$
root components (the maximal possible number) are allowed in every graph $\Gr$. 
In addition, there must be a consecutive subsequence of graphs $\Seqr{\rGST}{\rGST+4D-1} 
\subseteq \sigma$ where all root components are common\footnote{Recall that these common
root components are called $4D$-vertex-stable root components ($4D$-VSRCs)
in \cite{SWSBR15:NETYS}.} and ensure dynamic network diameter 
$D$. On the other hand, $\mathtt{MAJINF}(1)$ guarantees that the first $2D+1$-VSRC
that occurs in a run dominantly influences every subsequent $2D+1$-VRSC. This
ensures that a decision value possibly generated in an earlier $2D+1$-VRSC is 
duly propagated to every subsequent $2D+1$-VSRCs.
In the following \cref{cor:NETYS-vs-EStable}, we show that
$\mathtt{VSRC}(n, 4D) + \mathtt{MAJINF}(1)$
is stronger than $\EStable$. This implies that the
consensus algorithm from \cite{SWSBR15:NETYS} works also under $\EStable$.

\begin{theorem}
  \label{cor:NETYS-vs-EStable}
  Message adversary 
  $\mathtt{VSRC}(n, 4D) + \mathtt{MAJINF}(1)$
  is stronger than $\EStable$, i.e.,
$\mathtt{VSRC}(n, 4D) + \mathtt{MAJINF}(1)
\supseteq
\EStable$
\end{theorem}
\begin{proof}
  Since both adversaries guarantee the dynamic diameter $D$, it suffices
  to show that 
  $\mathtt{VSRC}(n, 4D) \supseteq \Liveness$
  and
  $\mathtt{MAJINF}(1) \supseteq \Safety(D)$
  both hold.
  
  $\mathtt{VSRC}(n, 4D) \supseteq \Liveness$:
  Take any feasible sequence $\sigma$ of $\Liveness$.
  By \cref{def:liveness}, there is some round $\rSR\geq\rGST$ from which
  on $\Seqr{\rSR}{\infty} \subseteq \sigma$ is $R$-single-rooted.
  But then also $\Seqr{\rSR}{\rSR+4D-1}$ is $R$-single-rooted
  and hence $\sigma \subseteq \mathtt{VSRC}(n, 4D)$.

  $\mathtt{MAJINF}(1) \supseteq \Safety(D)$:
  Pick an arbitrary feasible sequence $\sigma$ of $\Safety(D)$.
  If there is a subsequence $\SeqrI$ of $\sigma$ with
  common root $R$ consisting of $>2D$ rounds, then it follows 
  from \cref{def:safety} that there cannot be
  a subsequence $\SeqrJ$ of $\sigma$ with common root $R' \neq R$
  consisting of $>2D$ rounds, as $R$ and $R'$ both
  would need to be the single root of $\Seqr{\rSR}{\infty}$.
  Hence, $\sigma$ is trivially in $\mathtt{MAJINF}(1)$.
\end{proof}
}{}

\section{Termination time lower bound}

\iftoggle{TR}{It follows immediately from \cref{cor:NETYS-vs-EStable} that}
{As we show in \cref{cor:NETYS-vs-EStable} in the appendix,}
the gracefully degrading consensus algorithm from \cite{SWSBR15:NETYS} 
works also under $\EStable$. According to \cite[Lemma~5]{SWSBR15:NETYS},
it terminates at the end of round $\rSR + 4D$, i.e., has a termination
time of $4D+1$ rounds measured from the start of the stable period
(round $\rSR$).

From an applications perspective, fast termination is of course important.
An interesting question is hence whether the algorithm from \cite{SWSBR15:NETYS}
is optimal in this respect. The following \cref{thm:adversary-tightness} 
provides us with a lower bound of $2D$ for the termination time under \MA{} $\EStable$,
which proves that it is not: There is a substantial gap of $2D$ rounds.

\begin{theorem}\label{thm:adversary-tightness}
Solving consensus is impossible under message adversary $\EStable$
in round $\rSR+2D-1$.
\end{theorem}

\begin{figure}
  \centering
    \begin{subfigure}[b]{0.30\textwidth}
      \centering
      \begin{tikzpicture}
        \draw (0,0)     node[proc] (p1) {$p_1$};
        \draw (2,0)     node[proc] (p2) {$p_2$};
        \draw (-1,0)    node[proc] (p3) {$p_3$};
        \draw (-1,-0.7) node[proc] (p4) {$p_4$};
        \draw (1,-0.7)  node[proc] (p5) {$p_5$};

        \draw[dot] (p1) to (p3); 
        \draw[dot] (p1) to (p4); 
        \draw[dot] (p1) to (p5); 
        \draw[dot] (p5) to (p2); 
      \end{tikzpicture}
      \caption*{Round $1$ to $\infty$}
    \end{subfigure}
    \caption{Execution $\varepsilon$ of \cref{thm:adversary-tightness}, $n=5$, $D=2$}
  \label{fig:EStable-impossibility-eps1}
\end{figure}
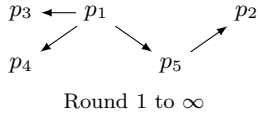

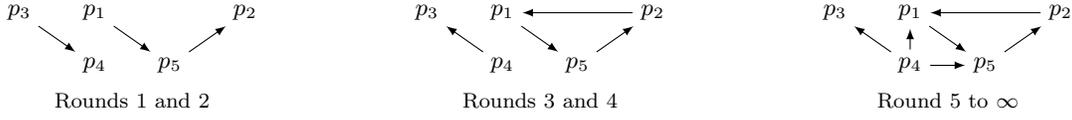
\begin{figure*}[h!]
  \centering
    \begin{subfigure}[b]{0.30\textwidth}
      \centering
      \begin{tikzpicture}
        \draw (0,0)    node[proc] (p1) {$p_1$};
        \draw (2,0)    node[proc] (p2) {$p_2$};
        \draw (-1,0)   node[proc] (p3) {$p_3$};
        \draw (0,-0.7) node[proc] (p4) {$p_4$};
        \draw (1,-0.7) node[proc] (p5) {$p_5$};

        \draw[dot] (p3) to (p4); 
        \draw[dot] (p1) to (p5); 
        \draw[dot] (p5) to (p2); 
      \end{tikzpicture}
      \caption*{Rounds $1$ and $2$}
    \end{subfigure}
    \begin{subfigure}[b]{0.30\textwidth}
      \centering
      \begin{tikzpicture}
        \draw (0,0)    node[proc] (p1) {$p_1$};
        \draw (2,0)    node[proc] (p2) {$p_2$};
        \draw (-1,0)   node[proc] (p3) {$p_3$};
        \draw (0,-0.7) node[proc] (p4) {$p_4$};
        \draw (1,-0.7) node[proc] (p5) {$p_5$};

        \draw[dot] (p4) to (p3); 
        \draw[dot] (p1) to (p5); 
        \draw[dot] (p5) to (p2); 
        \draw[dot] (p2) to (p1); 
      \end{tikzpicture}
      \caption*{Rounds $3$ and $4$}
    \end{subfigure}
    \begin{subfigure}[b]{0.30\textwidth}
      \centering
      \begin{tikzpicture}
        \draw (0,0)    node[proc] (p1) {$p_1$};
        \draw (2,0)    node[proc] (p2) {$p_2$};
        \draw (-1,0)   node[proc] (p3) {$p_3$};
        \draw (0,-0.7) node[proc] (p4) {$p_4$};
        \draw (1,-0.7) node[proc] (p5) {$p_5$};

        \draw[dot] (p4) to (p5); 
        \draw[dot] (p1) to (p5); 
        \draw[dot] (p4) to (p3); 
        \draw[dot] (p4) to (p1); 
        \draw[dot] (p5) to (p2); 
        \draw[dot] (p2) to (p1); 
      \end{tikzpicture}
      \caption*{Round $5$ to $\infty$}
    \end{subfigure}
    \caption{Execution $\varepsilon'$ of \cref{thm:adversary-tightness}, $n=5$, $D=2$}
  \label{fig:EStable-impossibility-eps2}
\end{figure*}

\begin{proof}
We will use a contradiction
proof based on the indistinguishablility of specifically constructed
admissible executions. Since the processes have no knowledge of 
$\Pi$ and $\abs{\Pi}$, we can w.l.o.g.\ assume that $n \geq 4$ and 
$D < n-2$. 

\def\C{\mathcal{C}}
Assume that an algorithm $\mathcal{A}$ exists that 
solves consensus under $\EStable$ by the end of round $\rSR+2D-1$.
Then, $\mathcal{A}$ must also solve consensus in the following execution
$\varepsilon$: In $\varepsilon$, all processes in $\Pi$ start with input value $0$,
and all graphs in $\Seqr{1}{\infty}$ are the same $\G$. The graph $\G$ is
single-rooted with $R=\{p_1\}$ and contains a chain $\C \subset \G$ consisting of $D+1$ 
processes $C \subseteq \Pi$ that starts in $p_1\in C$ and ends in $p_2\in C$. 
All remaining processes
are direct out-neighbors of $p_1$. \cref{fig:EStable-impossibility-eps1}
shows an example of the graph $\G$ used in $\varepsilon$ for $n=5$ and $D=2$. The execution is admissible because its
graph sequence is feasible for $\EStable$ with $\rSR = \rGST = 1$.
By validity and our termination time assumption, every process must 
hence have decided $0$ by the end of round
$\rSR+2D-1$ in $\varepsilon$.

We will now construct an execution $\varepsilon'$ of $\mathcal{A}$, 
where some process in $\Pi\setminus\{p_1,p_2\}$ eventually decides 1
albeit the state 
$\state{p_2}{\rSR+2D-1}$ of process $p_2$ at the end of 
round $\rSR+2D-1$ is the same as in $\varepsilon$.
Thus, $\varepsilon$ and $\varepsilon'$ are
indistinguishable for process $p_2$ until $\rSR+2D-1$.
An example of the graph sequence used in $\varepsilon'$ 
for $n=5$ and $D=2$ is shown in \cref{fig:EStable-impossibility-eps2}.

In $\varepsilon'$, let two processes $\{p_3,p_4\}$ in 
$\Pi\setminus C$ have initial value $1$ and 
all remaining ones have initial value $0$.
The identical graph $\G'$ used in $\Seqr{1}{D}$ consist of the very
same chain $\C$ as in $\G$, and a single edge $(p_3,p_4)$.
Note that $\G'$ contains two root components, namely
$R_1=\set{p_1}$ and $R_2=\set{p_3}$.
The identical graph $\G''$ used in $\Seqr{D+1}{2D}$ consist of the
chain $\C$, an additional edge $p_2$ to $p_1$, and an
edge $(p_4,p_3)$. Again, $\G''$ contains two
root components,  $R_1=C$ and $R_2=\set{p_4}$. Finally, the graph
$\G'''$ used in $\Seqr{2D+1}{\infty}$ is $\G''$ augmented
by two edges connecting $p_4$ to two different
process in $C$.
Note that it contains a single root $R=\set{p_4}$ and guarantees a
dynamic diameter of (at most) $D$.

Clearly, $\varepsilon'$ is an admissible execution for $\EStable$: It adheres to $\Liveness$
for $\rSR = D+1$, when $\set{p_4}$ becomes a forever common root that becomes
single forever starting with round $2D+1$. It is also feasible for 
$\Safety(D)$, as the only graph sequence 
that contains a common root for more than $D$ rounds, namely, the final
one $\Seqr{2D+1}{\infty}$, is single-rooted.

For $p_2$, the executions $\varepsilon$ and $\varepsilon'$
are indistinguishable for the first $2D$ rounds, because
by the end of round $2D$, $p_2$ cannot have learned of the existence 
of the edge $(p_2 \ra p_1)$ that distinguishes the root components $R$ and $R_1$
involving $p_1$ in $\G$ and $\G''$, respectively:
It takes at least $D$ rounds for any information, sent by $p_1$, to be forwarded along $\C$
to $p_2$, and $p_1$ cannot have learned about the existence of this edge 
before round $D+1$.
It hence follows that $p_2$ decides $0$ in round $2D$ also in $\varepsilon'$,
as it does so in $\varepsilon$.

In $\varepsilon'$, by validity and the assumed correctness of
$\mathcal{A}$, however, all processes must eventually decide $1$ to solve 
consensus: The only input value that $p_4$ ever gets to know
throughout the entire execution is 1. The same is true in the execution
$\varepsilon''$, which is identical to $\varepsilon'$ except that
the input value of all processes is 1. Clearly, $p_4$ must decide 1
in $\varepsilon''$ and, hence, also in $\varepsilon'$. This provides
the required contradiction and completes our proof.

Above, we have shown the impossibility for the case where $\rSR=1$ (which
would already be sufficient for the claim of \cref{thm:adversary-tightness}).
Actually, it is not hard extend the proof for general $\rSR$, by simply prefixing
$\varepsilon$ and $\varepsilon'$ with the following graph sequence $\pi$:
In every round $\leq \rSR$ of $\pi$, the graphs alternate between $\G'$ and $\G''$, such that
the graph in the last round of $\pi$ is $\G''$.
The resulting prefixed executions obviously still adhere to the
\MA{} $\EStable$ and
are indistinguishable from their respective prefixed
counterparts for processes $p_2$ and $p_4$.
\end{proof}

We will show in the next section that the lower bound established in
\cref{thm:adversary-tightness} is tight, by providing a matching algorithm.

\section{A fast consensus algorithm} \label{sec:algorithm}

\newcommand\lock[1]{\mathtt{lock}_{#1}}
\newcommand\round{\mathtt{r}}
\newcommand\lockround{\mathtt{lockRound}}
\newcommand\counter{\mathtt{count}}
\newcommand\lockval{\ell}
\newcommand\Root{R}

We now present our consensus algorithm for the message adversary $\EStable$,
which also works correctly under the generalized $\AltEStable$ 
that will be introduced in \cref{sec:refinements}.
The algorithm is based on the fact that, from the messages a node receives, it
can reconstruct a faithful under-approximation of (the relevant
part of) the communication graph of every round, albeit with delay $D$.

\begin{figure*}[ht!]
 \centering
 \scalebox{.9}{
 \begin{tikzpicture}[
   every node/.style={inner sep=3pt},
   cond/.style={draw, rounded corners, align=left},
   branch/.style={draw, align=left},
   >=latex,shorten <=0pt,shorten >=0pt,                                          
   link/.style={->,>=stealth'}]
   \def\yorigin{-0.4}
   \def\dx{8}
   \def\dy{-2}
   \newcommand\yes[2]{\draw[link] (#1) -- node[auto] {yes} (#2)}
   \newcommand\no[2]{\draw[link] (#1) -- node[auto, near start] {no} (#2)}
   \newcommand\notip[2]{\draw[] (#1) -- node[auto, near start] {no} (#2)}

   \node[draw, densely dotted] (start) at (0,\yorigin) {start of round $m\geq D+1$ computation}; 
   \node[draw, densely dotted] (end) at (\dx,\yorigin) {end of round $m$ computation}; 
   \node[cond] (c1) at (0,\dy)
      {\textbf{c1:}\\
      Check if $\ApproxGp{m-D}$ contains exactly one root $R$ such that\\
	  for every $q\in R$ there is
      a round $m'>m-D$ s.t.\ \\
      there is an outgoing edge from  $q$ in $\ApproxGp{m'}$.\\
                      };
   \node[branch] (b1) at (0,2.2*\dy)
      {\textbf{b1:}\\
	  Find a round $a$ such that $R$ is a maximal common root of $(\ApproxGp{r})_{r=a}^{b}$ \\
          and $(m-D)\in [a,b]$. Set $\lock{}[p][m]$ to the maximum
	  $\lock{}[q][a]$\\ of all processes $q \in R$.};
   \node[cond] (c2) at (0, 3.2*\dy)
   {\textbf{c2:}\\  Does there exist a $R'$-single-rooted sequence $(\ApproxGp{r})_{r =a'}^{b'}$ with $b'-a'+1 > D$?};
   \node[cond] (c3) at (0, 4.0*\dy)
      {\textbf{c3:}\\ For every process $q$ of $R'$, is there some round $\gamma>b'$ s.t.\\
      there is an outgoing edge from $q$ in $\ApproxGp{\gamma}$?};
   \node[branch] (b3) at (0,4.9*\dy)
   {\textbf{b3:}\\ If not yet decided, find a round $a''$ s.t.\ $R'$ is maximum common root
   of
   $(\ApproxGp{r})_{r=a''}^{b''}$ and\\ $[a',b'] \subseteq [a'', b'']$ and 
   decide on the maximum $\lock{}[q][a'']$
   of all processes $q \in R'$.
    };
   
   \def\Yaa{2.8*\dy}
   \def\Xa{\dx-2.5, 1*\dy}
   \def\Xaa{\dx-2.5, \Yaa}
   \def\Xb{\dx, 3.2*\dy}
   \def\Xc{\dx, 4.0*\dy}
   \def\Xd{\dx, 4.9*\dy}
   
   \draw[link] (start) -- (c1);
   \notip{c1}{\Xa};
   \draw[link,-] (\Xa) -- (\Xaa);
   \draw[link,-] (\Xaa) -- (0,\Yaa);
   \yes{c1}{b1};
   \draw[link] (b1) -- (c2);
   \notip{c2}{\Xb};
   \yes{c2}{c3};
   \notip{c3}{\Xc};
   \yes{c3}{b3};
   \draw (b3) -- (\Xd);
   \draw[link] (\Xd) -- (end);
 \end{tikzpicture}}

 \caption{Round $m\geq D+1$ core computation step of our consensus algorithm for 
process $p$. $\ApproxGp{r} = \ApproxG{r}{p}{m}$ denotes
 $p$'s round $m$ view of $\Gr$ provided by the network approximation
algorithm. $\lock{}[q][r]$ denotes $\lock{p}^m[q][r]$, where
 $\lock{}[p][m]$ represents $p$'s proposal value for the next round $m+1$.}
 \label{fig:algorithm}
\end{figure*}
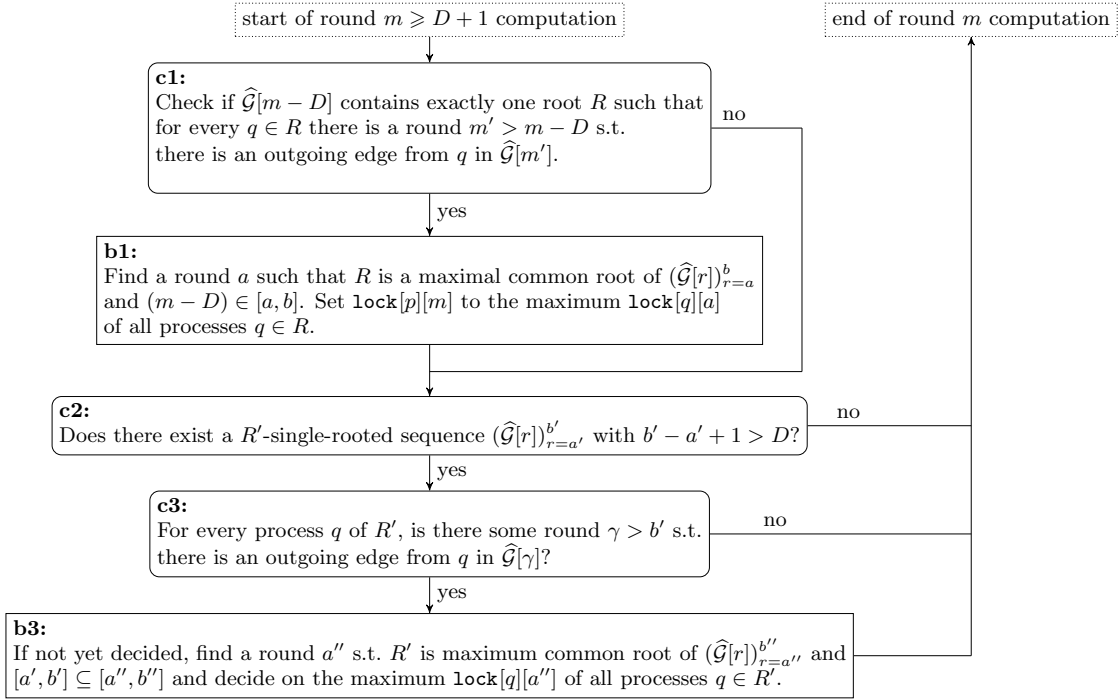

The algorithm stated in \cref{fig:algorithm} works as follows:
Every process $p$ maintains an array $\ApproxG{r}{p}{}$
that holds the \emph{graph approximation} of $\Gr$, and a matrix $\lock{p}[q][r]$
that holds the history of a special value, the \emph{lock-value}, for
every known process $q$ and every round $r$. $\ApproxG{r}{p}{m}$ and $\lock{p}^{m}[q][r]$ denote the
content of the respective array entry at the end of round $m$ as usual. 
The first entries of these arrays are initialized to 
the singleton-graph $\ApproxG{0}{p}{0} = (\{p\},\set{})$ 
resp.\ to $\lock{p}^0[p][0]:=x_p$, the input value of $p$, and to
$\lock{p}^0[q][0]:=\bot$ for every $q\neq p$. Note that  
$\lock{p}[p][m-1]$ can be viewed as $p$'s \emph{proposal value} for round $m$.
Every process broadcasts
$\ApproxG{r}{p}{m-1}$ and $\lock{p}^{m-1}[q][r]$ in round $m\geq 1$, and
updates $\ApproxG{r}{p}{m}$ and $\lock{p}^{m}[q][r]$, by fusing the information
contained in the messages received in round $m$ in a per-round fashion (as detailed below), 
before executing the
round $m$ \emph{core computation} (we will omit the attribute
core in the sequel if no ambiguity arises) of the algorithm. 
Note that the round $m$ core computation for $m\in\set{1,\dots,D}$ 
is empty.

In the computation of some round $\tau$, 
$p$ will eventually
decide on the maximum $\lock{p}[q][a]$ value for all
$q \in R$, where $R$ is a common root of some sequence
$\Seqr{a}{a+D}$ but not of $\Seqr{a-1}{a+D-1}$, as detected locally
in $\ApproxG{*}{p}{\tau}$.
Note carefully that $\tau$ may be different for processes other than $p$.

Two mechanisms are central to the algorithm for accomplishing this:
First, any process $p$ that, in its round $m$ computation, locally 
detects a single 
root component $R$ in $\ApproxG{m-D}{p}{m}$ will ``lock'' it, 
i.e., assign
the maximum value of $\lock{p}^m[q][m-D]$ for any $q \in R$  to $\lock{p}^m[p][m]$.
Second, if process $p$ detects in round $\tau$ that a graph sequence had a
common root $R'$ for at least $D+1$ rounds in its graph approximation, 
starting in round $a$, $p$ will
decide, i.e., set $y_p$ to the maximum of $\lock{p}^{\tau}[q][a]$ among all $q\in
R'$.

Informally, the reason why this algorithm works
is the following: 
From detecting an $R$-single-rooted sequence of length $\geq D+1$,
$p$ can infer, by the $\Safety(D)$ property of our \MA,
that the entire system is about to lock $p$'s decision value.
Moreover, by exploiting the information propagation guarantee given
by \cref{cor:end-to-end}, we can be sure that, after $p$'s 
decision in round $\tau$,  every other process $q$ decides
(in some round $\tau'\geq \tau$) on the very same value: Under $\EStable$,
it decides because the root that triggered the decision of 
$p$ is the \FAES-common root; under $\AltEStable$, $q$
decides on the same value because it will never assign 
a value different from $\lock{p}[p][\tau]$ to
$\lock{q}[x][\tau'']$ for any $\tau'' \geq \tau'$ and any known process $x$.
Finally, termination is guaranteed since every $p$ will eventually 
find an $R$-single-rooted sequence of
duration at least $D+1$ because of $\Liveness$.

\subsection*{Graph approximation and lock maintenance}

Our algorithm relies on a simple mechanism for maintaining the graph approximation
$\ApproxG{r}{p}{}$ and the array of lock values $\lock{p}[q][r]$ at every process $p$:
In every round, each process $p$ broadcasts its current 
$\ApproxG{*}{p}{}$ and $\lock{p}[*][*]$ and updates all entries with new information
possibly obtained in the received approximations from other processes. 
In more detail, an edge $(q \ra q')$ will be present in $\ApproxG{r}{p}{m}$ 
at the end of round $m\geq r$ if either $p = q'$
and $p$ received a message from $q$ in round $r$, or if
$p$ received $\ApproxG{r}{q''}{r''}$ for $m \geq r'' \geq r$ 
from some process $q''$ and $(q \ra q') \in \ApproxG{r}{q''}{r''}$.
Similarly, $\lock{p}^{m}[q][r]$ for $r< m$ is updated to
$\lock{q'}[q][r] \neq \bot$ whenever such an entry is 
received from any process $q'$; the entry
$\lock{p}^{m}[q][m]$ for the current round $m$ is initialized to 
$\lock{p}[p][m]:=\lock{p}[p][m-1]$ for $q=p$ and 
to $\lock{p}[q][m]:=\bot$ for every $q\neq p$. 

Note carefully that we assume that the round $m$ computation of the approximation algorithm 
is executed \emph{before} the round $m$ core computing step at every process.
Therefore, the round $m$ approximation $\ApproxG{*}{p}{m}$ is already 
available \emph{before} the core computing step of round $m$ at process $p$ is
executed.

We do not provide further details of the implementation of this graph approximation
here; a fitting algorithm, along with its correctness proof, can be found 
in \cite{SWSBR15:NETYS,BRS12:sirocco}. We remark, though, that the 
full-information approach of the above implementation incurs sending 
and storing a large amount of redundant information. Comments related to
a more efficient implementation are provided in
\iftoggle{TR}{\cref{sec:refinements}}{the full version \cite{SSW15:TR} of our paper}.

The crucial property guaranteed by our graph approximation is that processes
under-approximate the actual communication graph, i.e., that they do not 
fabricate edges in their approximation. Using our notion of causal past,
it is not difficult to prove the following assertion about edges that are 
guaranteed to exist in the graph approximation:

\begin{lemma}
  In a full-information graph approximation protocol,
  $\infl{q}{r}{p}{r'}$ holds for $r'>r$ $\Leftrightarrow$ there exists a process $q'$ s.t.\
  $(q \ra q') \in \ApproxG{r''}{p}{r'}$ for some $r'' \in (r, r']$.
  \label{cor:late-edge}
\end{lemma}

\begin{proof}
  ``$\Rightarrow$''-direction:
  If $p=q$, the claim trivially holds because every communication graph
  contains the self-loop $(p \ra p)$.
  For $p \neq q$, since we assume $\infl{q}{r}{p}{r'}$, by
  \cref{def:causal-influence}, there exists a round $r'' > r$, such that
  $\exists \infl{q'}{r''}{p}{r'}$ with $(q \ra q') \in \G^{r''}$.
  Therefore, $p$ must have received the round $r''$ state
  of $q'$ and hence learned about the edge $(q \ra q')$, by round $r'$.
  In other words, $(q \ra q') \in \ApproxG{r''}{p}{r'}$, as claimed.

  ``$\Leftarrow$''-direction:
  Since we assume a full-information protocol, $p$ knowing part of the state of
  another process $q'$ implies that $p$ knows the entire state of $q'$.
  Hence,
  if $(q \ra q') \in \ApproxG{r''}{p}{r'}$, $p$ knows the state of $q'$ of
  round $r''$. 
  Thus $\infl{q'}{r''}{p}{r'}$ with 
  $r'' \leq r'$.
  From \cref{cor:transitive-influence}, it follows that
  $\infl{q}{r''-1}{p}{r'}$, which implies $\infl{q}{r}{p}{r'}$ because $r \leq r''$.
\end{proof}

We now present a more abstract view on this mechanism of approximating
the communication graph.
First, we answer which state information a process needs in order to
reliably detect which roots are present in the actual
communication graph.

\begin{lemma} 
  Let $R \in \roots(\Gr)$ and let there be some process $p$ and round $r'$ such that
  $\setinfl{R}{r}{p}{r'}$.
  In a full-information graph approximation protocol, $R \in \roots(\ApproxG{r}{p}{r'})$.
  Furthermore,
  there exists a process $q'$ s.t.\
  $(q \ra q') \in \ApproxG{r''}{p}{r'}$ for some $r < r'' \leq r'$.
  \label{lem:root-detection}
\end{lemma}

\begin{proof}
  Since $\setinfl{R}{r}{p}{r'}$, according to \cref{cor:transitive-influence},
  by the end of round $r'$,
  $p$ has received the round $r$ state
  $\state{q}{r}$ of all processes $q \in R$.
  In particular, $p$ has received all round $r$ in-edges of every process $q$.
  Hence, $R$ is a strongly connected component of $\ApproxG{r}{p}{r'}$
  and there are no processes $q' \in \Pi \setminus R$ s.t.\
  $(q' \ra q) \in \ApproxG{r}{p}{r'}$.
  But then, $R \in \roots(\ApproxG{r}{p}{r'})$, as asserted.
  The presence of $(q \ra q')$ in $\ApproxG{r''}{p}{r'}$ follows directly from
  \cref{cor:late-edge}.
\end{proof}

We conclude our considerations regarding the graph approximation 
by looking at what is sufficient from an algorithmic
point of view for a process $p$ to faithfully determine the root components
in some communication graph.
In the case where a root component $R \in \roots(\Gr)$ has size $|R| > 1$,
we note that as soon as a process $p$ knows, in some round $r'$, at least one in-edge
$(q' \ra q) \in \ApproxG{r}{p}{r'}$
for each $q \in R$, then $p$ knows $\state{q}{r}$ and hence all in-edges of $q$.
Consequently, it can reliably deduce that indeed $R \in \roots(\Gr)$.

In the case where $|R|=\abs{\set{q}}=1$, if $p$ has no edge
$(q' \ra q) \in \ApproxG{r}{p}{r'}$, this is \emph{not} sufficient for
concluding that $\set{q} \in \roots(\Gr)$:
Process $p$ seeing no in-edge to a process $q$ in the local graph
approximation $\ApproxG{r}{p}{r'}$ happens naturally if
$\infl{q}{r-1}{p}{r'}$
and
$\ninfl{q}{r}{p}{r'}$, i.e., when
the last message $p$ received from $q$ was sent at the beginning
of round $r$.
In order to overcome this issue, process $p$ must somehow ascertain
that it
already received the state $\state{q}{r}$ of process $q$ in round $r$.
In particular, process $p$ can deduce this directly from its graph approximation
as soon as it observed some
outgoing edge from $q$ in a round strictly after $r$.

Let us state this more formally in the following lemma.

\begin{lemma}\label{lem:root-consistency}
  Consider a full-information graph approximation protocol.
  Let $R \in \roots(\ApproxG{r}{p}{r'})$ for $r'>r$, and let, for all processes $q \in R$,
there be a process $q'$ and a round $r'' \in (r, r']$, such that
  $(q \ra q') \in \ApproxG{r''}{p}{r'}$.
  Then, $R \in \roots(\Gr)$,
  and $\setinfl{R}{r}{p}{r'}$.
\end{lemma}

\begin{proof}
  By contradiction.
  Assume that $R \in \roots(\ApproxG{r}{p}{r'})$,
  $\forall q \in R \, \exists q' \in \Pi, r'' \in (r, r'] \colon (q \ra q') \in \ApproxG{r''}{p}{r'}$
  and $R \notin \roots(\Gr)$.
  Because of the latter, there exist some processes $q \in R$ and $q'' \notin R$ with
  $(q'' \ra q) \in \Gr$.
  By the presence of the edge $(q \ra q')$ in $\ApproxG{r''}{p}{r'}$ and
  \cref{cor:late-edge},
  we have
  $\setinfl{R}{r}{p}{r'}$.
  But then, by the assumption that $(q'' \ra q) \in \Gr$,
  it must also hold that $(q'' \ra q) \in \ApproxG{r}{p}{r'}$.
  This, however, contradicts that $R \in \roots(\ApproxG{r}{p}{r'})$.
\end{proof}

Finally, the way how the lock arrays are maintained by our algorithm
implies the following simple results: 

\begin{corollary}\label{cor:stateimpllock}
  If $r'>r$, then $\infl{q}{r}{p}{r'}$ implies that also
$\lock{p}^{r'}[q][r'']= \lock{q}^{r''}[q][r'']$
  for all rounds $r'' \leq r$.
\end{corollary}

\begin{lemma}\label{lem:stateimpllockown}
Let $m$ be a round reached by process $p$ in the execution. Then,
$\lock{p}^{m}[p][r] \neq \bot$ for all $0\leq r \leq m$.
\end{lemma}

\begin{proof}
Since $\lock{p}^0[p][0]=x_p$, it follows from the update rule $\lock{p}[p][m]:=\lock{p}[p][m-1]$
that $\lock{p}[p][m] \neq \bot$ for all reached rounds $m$, provided that the core algorithm never 
assigns $\bot$ in b1. Since the latter can only assign the maximum of $\lock{p}[q][a]$ for all
$q\in R$ from some earlier round $a \leq m-D < m$, the statement of our lemma
follows from a trivial induction based on \cref{cor:stateimpllock}, provided we can guarantee
$\infl{q}{a}{p}{m}$. The latter follows immediately from c1 in conjunction with \cref{lem:root-consistency},
however.
\end{proof}

\subsection*{Correctness proof} \label{sec:correctness}

Before proving the correctness of the algorithm given in \cref{fig:algorithm}
(\cref{thm:correctness} below),
we first establish two technical lemmas: \cref{lem:termination} 
reveals that our
algorithm terminates for every message adversary $\MAT$ that guarantees 
certain properties (without guaranteeing agreement, though). The complementary
\cref{lem:agreement} shows that our algorithm ensures agreement (without
guaranteeing termination, though) for
every message adversary $\MAA$ that guarantees certain other properties. 
\cref{thm:correctness} will then follow from the fact that
$\EStable \subseteq \MAT \cap \MAA$.

\begin{lemma}
  The algorithm terminates by the end of round $\tau$ under any message adversary $\MAT$
  that guarantees dynamic diameter $D$ in conjunction with
  the following properties: For every $\sigma \in \MAT$,
  \begin{compactitem}
  \item there is an $R$-single-rooted sequence $\Seqr{\alpha}{\beta} \in \sigma$ with
    $\beta-\alpha+1 > D$.
  \item there is a round $\tau$ such that $\setinfl{R}{\beta}{p}{\tau}$, 
for all $p \in \Pi$.
  \end{compactitem}
  \label{lem:termination}
\end{lemma}

\begin{proof}
  We show that if process $p$ has not decided before round $\tau$, it will do so in
  round $\tau$.
  By round $\tau$, every process $p \in \Pi$ received $\state{q}{\beta}$ for all $q \in R$
  by the assumption that $\setinfl{R}{\beta}{p}{\tau}$.
  Hence, by \cref{lem:root-detection} and \cref{lem:root-consistency}, for every
  $p \in \Pi$, it holds that $R$ is the single root of $\roots(\ApproxG{\beta}{p}{\tau})$.
  Furthermore, by \cref{cor:transitive-influence}, $R$ is in fact
  the single root of $\roots(\ApproxG{r}{p}{\tau})$ for any $r \in [\alpha, \beta]$.
  Therefore, process $p$ will pass the check c2 in round $\tau$.

  In addition, by the assumption that $\setinfl{R}{\beta}{p}{\tau}$ and \cref{lem:root-detection},
  for every $q \in R$, there exists a round $\beta' \in (\beta, \tau]$, s.t.\
  $(q \ra q') \in \ApproxG{\beta'}{p}{\tau}$ for some process $q'$.
  Therefore, process $p$ will pass the check c3 in round $\tau$ and decide.
\end{proof}

\cref{lem:agreement} below shows that, under message adversaries that 
guarantee a $\ECS(D+1)$-common root according to \cref{def:ECS-common-root}, 
the algorithm from \cref{fig:algorithm} 
satisfies agreement.
\iftoggle{TR}{}{Due to space constraints, the proof had to be relegated to the appendix.
}

\begin{definition}\label{def:ECS-common-root}
We say that a graph sequence $\Seqr{\alpha}{\alpha+d}$ has a \emph{$\ECS(x+1)$-common root}
(``embedded $x+1$-consecutive single common root'') $R$, if
(i) $\Seqr{\alpha}{\alpha+d}$ has a common root $R$ and
(ii)  $\Seqr{\alpha'}{\alpha'+x}\subseteq \Seqr{\alpha}{\alpha+d}$ 
has a single root $R$.
\end{definition}

\begin{lemma}
  Let $\MAA$ be a \MA{} that guarantees, for every $\sigma \in \MAA$, 
a dynamic diameter $D$ in conjunction with the property that
the first subsequence $\Seqr{\alpha}{\beta} \subseteq \sigma$ with a maximum
common root $R$ and $\beta-\alpha+1 > D$ has a $\ECS(D+1)$-common root.
 Under $\MAA$, if two or more processes decide in our algorithm, then they 
decide on the same value $\neq \bot$.
  \label{lem:safety}
  \label{lem:agreement}
\end{lemma}

\iftoggle{TR}{
  \begin{proof}%[\iftoggle{TR}{}{of \cref{lem:agreement}}]
Let $\alpha'$ and $\beta'$, with $\beta'-\alpha'+1>D$, delimit the maximal 
period where $R$ is single-rooted, as predicted by \cref{def:ECS-common-root}.

Setting $\lambda = \max_{q \in R} \lock{q}^{\alpha}[q][\alpha]$,
we show that if an arbitrary process $p$ decides in round $\tau$, 
it decides on $\lambda$ and $\lambda\neq \bot$.
  Assume that $p$ decides in some round $\tau$.
  It follows from c2 and c3 that $p$ detects in round $\tau$
  that $R'$ is the single root of $(\ApproxG{r}{p}{\tau})_{r=a'}^{b'}$
  with $b' - a' + 1 > D$,
  and that, for every $q\in R'$, there is a round $\gamma > b'$ where there
  is an edge $(q, q')$ in $\ApproxG{\gamma}{p}{\tau}$ for some process $q' \in \Pi$.
  By \cref{lem:root-consistency}, we have that $R' \in \roots(\Gr)$ for all
  $r \in [a', b']$, and $\setinfl{R'}{b'}{p}{\tau}$. Thus, \cref{cor:stateimpllock} in conjunction
with \cref{lem:stateimpllockown} confirm that indeed $\lambda\neq \bot$.
  We distinguish two cases:

  \emph{Case 1.} 
  $[a', b'] \subseteq [\alpha, \beta]$:
  From the definition of $\MAA$, in combination with the fact that
  $b'-a'+1 > D$, it follows that $R' = R$:
  if this was not the case, then either $\Seqr{\alpha}{\beta}$ would not
  be the first sequence of its kind or $\Seqr{\alpha'}{\beta'}$ would
  not be $R$-single-rooted.

  By b3, $p$ will decide on the maximum of $\lock{p}[q][a'']$, where
  $a''$ is a round such that $(\ApproxG{r}{p}{\tau})_{r=a''}^{b''}$
  has a maximum common root $R$,
  $[a'', b''] \supseteq [a', b']$,
  and $q \in R$.
  Hence, since $\setinfl{R}{b'}{p}{\tau}$ and $\alpha < b'$, it follows
  from \cref{cor:transitive-influence} that $\setinfl{R}{\alpha}{p}{\tau}$. 
  Thus, by \cref{lem:root-detection}, we have $a'' = \alpha$.
  According to \cref{cor:transitive-influence} in conjunction with
  \cref{cor:stateimpllock}, it follows that $p$ indeed decides on $\lambda$.

  \emph{Case 2.} $[a', b'] \nsubseteq [\alpha, \beta]$:
  First, observe that $a' > \beta'$:
  If $a' \leq \beta'$ then, because $\Seqr{\alpha}{\beta}$ is the first sequence of its kind,
  we have that $a' \geq \alpha$.
  Thus, since $\G^{\beta'}$ is $R$-single-rooted,
  $R' = R$, and hence $[a', b'] \nsubseteq [\alpha, \beta]$ is a contradiction to the
  assumption that $R$ is maximal common in
  $\Seqr{\alpha}{\beta}$.
  
  It follows from this observation and b3 that $p$ decides on the 
maximum value of $\lock{p}[q][a'']$ for $q \in R'$, where $a'' > \beta'$.  
  Thus, to conclude our proof, it suffices to show that
$\lock{p}^{r}[p][r] = \lambda$ for all rounds $r > \beta'$ and
all processes $p \in \Pi$.
  
  Since $\Seqr{\beta'-D}{\beta'}$ is $R$-single-rooted, it follows
from \cref{def:dynamic-diameter,lem:root-detection} that in round $\beta'$
every process $p$ sets $\lock{p}^{\beta'}[p][\beta']$ 
to $\lambda$ via b1. Moreover,
if a process assigns a value to $\lock{p}[p][m]$ during some round $m \in
  (\beta', \beta'+D]$ via b1 later on,
  it follows from the single-rootedness of $\Seqr{\beta'-D}{\beta'}$ and
  \cref{lem:root-consistency} that the assigned value is also $\lambda$.

  For $\ell\geq \beta'+D$, we show by induction 
on $\ell$ that $\lambda$ is assigned
to $\lock{p}[p][m]$ (if there is any assignment at all), 
in round $m$, for all $m\in [\beta',\ell]$ and all processes $p$. 
The induction basis is $\ell=\beta'+D$, for which the claim
has been established already. For the induction step, assume that
the claim holds for the interval $[\beta',\ell]$ and all $p$.
If no process $p$ changes its lock value in b1
  during the core round $\ell+1$ computation, i.e.,
  $\lock{p}^{\ell}[p][\ell]=\lock{p}^{\ell+1}[p][\ell+1]$,
  then the claim follows immediately from
  the induction hypothesis.
  Thus, assume that $\lambda=\lock{p}^{\ell}[p][\ell]\neq \lock{p}^{\ell+1}[p][\ell+1]$.
  This means that $p$ has successfully passed c1 and hence, by \cref{lem:root-consistency},
  that there is a root $R'' \in \roots(\G^{\ell+1-D})$ with
  $\setinfl{R''}{\ell+1-D}{p}{\ell+1}$.
  If $R''=R$ is a maximal common root of $\Seqr{\alpha}{\beta}$,
  by \cref{cor:transitive-influence}, it follows from the definition of $\lambda$ and
  \cref{cor:stateimpllock} that $p$ assigns
  $\lock{p}^{\ell+1}[p][\ell+1] := \lambda$.
  Therefore, assume that this is not the case, i.e., $R''\neq R$.
  Still, $R''$ must be a maximal common root in $\Seqr{\alpha''}{\beta''}$ for
  some $\alpha'' > \beta'$ with $\alpha'' \leq \ell+1-D$.
  By the induction hypothesis, $\lock{q}^{\ell+1-D}[q][r] = \lambda$ for every process
  $q$ of $R''$ and round $r \in [\beta', \ell]$ and
  so, in particular, $\lock{q}^{\ell+1-D}[q][\alpha''] = \lambda$.
  It follows from \cref{cor:stateimpllock} and $\setinfl{R''}{\ell+1-D}{p}{\ell+1}$
  that for all processes $q \in R''$, we have $\lock{p}^{\ell+1}[q][\alpha''] = \lambda$.
  Therefore, since, by b1, $p$ chooses its new value for $\lock{p}^{\ell+1}[p][\ell+1]$ as the maximum
  of the entries $\lock{p}^{\ell+1}[q][\alpha'']$,
  it assigns $\lock{p}^{\ell+1}[p][\ell+1] :=
  \lambda$.
\end{proof}
}{}

\begin{theorem} \label{thm:correctness}
The algorithm from \cref{fig:algorithm} solves consensus by round $\rSR + 2D$ under 
\MA{} $\EStable$.
\end{theorem}

\begin{proof}
According to b3, a process $p$ can decide only on a value in $\lock{p}^m[p][*]$
in some round $m$. By \cref{lem:stateimpllockown}, this value must be $\neq \bot$.
Since $\lock{q}[q][0]$ is initialized to $x_q$ for any process $q$, 
  and the only assignments $\neq \bot$ to any $\lock{q}$ entry
  are $\lock{q'}$ entries of other processes, validity follows.

  For agreement, recall that $\Safety(D)$ guarantees that the first sequence $\SeqrI$
  with a common root $R$ and $|I|>D$ must be the \FAES-common root.
  Hence, agreement follows from \cref{lem:agreement}.
 
  For termination, recall that $\Liveness$ guarantees the existence of some 
round $\rSR\geq \rGST$ such that $\Seqr{\rSR}{\infty}$ is $R$-single-rooted.
  This implies that the sequence $\Seqr{\rSR}{\rSR+D}$ is $R$-single-rooted 
  and, by \cref{def:dynamic-diameter}, $\setinfl{R}{\rSR+D}{p}{\rSR+2D}$.
  \cref{lem:termination} thus implies
  termination by round $\rSR+2D$.
\end{proof}

\section{Generalized stabilizing message adversary}\label{sec:refinements}

%\subsubsection*{A stronger stabilizing message adversary}

The simple message adversary introduced in \cref{sec:adversary}
may be criticized due to the fact that the first root component 
$R$ that is common in at least $D+1$ consecutive rounds must already 
be the \FAES-common root that persists forever after. In this 
section, we will considerably relax this assumption, which is
convenient for analysis and comparison purposes but 
maybe unrealistic in practice.

In the following \cref{def:alt-liveness}, we
start with a significantly relaxed variant $\AltLiveness(x)$ 
of $\Liveness$ from \cref{def:liveness}: Instead of requesting an infinitely
stable \FAES-common root $R$, we only require $R$ to be (i) a $\ECS(x+1)$-common 
root that starts at $\rGST$ and becomes single at $\rSR\geq\rGST$, 
and (ii) to re-appear as a single root in at least $D$ not necessarily 
consecutive later round graphs $\G^{r_1},\dots,\G^{r_D}$. 
Note that, according to \cref{def:dynamic-diameter}, the latter condition ensures 
$\setinfl{R}{\rSR+x}{p}{r_D}$ for all $p \in \Pi$ if $\AltLiveness(x)$ adheres
to the dynamic diameter $D$. 

\begin{definition}\label{def:alt-liveness}
%We call a root component $R$ starting in some round $a$ in $\sigma=\SeqInf$, 
%which (i) is maximal common in a subsequence
%$\Seqr{a}{a+d}$ of some $d+1$ consecutive rounds,
%and (ii) is single in a subsequence 
%$\Seqr{a'}{a'+x}\subseteq \Seqr{a}{a+d}$ of $x+1$ consecutive rounds
%a \emph{$\ECS(x+1)$-common root} (``embedded $x+1$-consecutive single 
%common root'').
%We say that a sequence $\Seqr{a}{a+d}$ has a \emph{$\ECS(x+1)$-common root}
%(``embedded $x+1$-consecutive single common root'') $R$ if
%(i) $\Seqr{a}{a+d}$ has maximal common root $R$ and
%(ii)  $\Seqr{a'}{a'+x}\subseteq \Seqr{a}{a+d}$ has single root $R$.

Every communication graph sequence $\sigma \in \AltLiveness(x)$ 
%For any run $\sigma \in \AltLiveness_D(X)$, 
contains a subsequence $\Seqr{\alpha}{\alpha+d}$, which has a $\ECS(x+1)$-common root
$R$; let $\rGST=\alpha$ be its starting round and $\rSR=\alpha'$ be the
time when it becomes single.
Furthermore, there are
at least $D$, not necessarily consecutive, $R$-single rooted round graphs
$\G^{r_1}, \ldots, \G^{r_D}$ with $\rSR+x<r_1 < \dots < r_D$ in $\sigma$.
%Moreover, $R$ re-appears as a single root in 
%at least $D$ not-necessarily consecutive round graphs $\G^{r_1}, \ldots, \G^{r_D}$ 
%with $\rSR+x<r_1 < \dots < r_D$.
%\begin{equation*}
%\abs{ \set{ \Gr \in \sigma \setminus \Seqr{1}{\rSR+x} \mid \roots(\Gr) = \set{R} }} \geq D.
%\end{equation*}
\end{definition}

Moreover, we also relax the $\Safety(x)$ condition in \cref{def:safety} accordingly:
We only require that the first root component $R$ that is common for at least $x+1$
consecutive rounds in a graph sequence $\sigma=\SeqInf$ is a $\ECS(x+1)$-common root:

\begin{definition}
For every $\sigma \in \AltSafety(x)$, it holds that the
%Let $\sigma \in \AltSafety_D(X)$.
%Then $\sigma$ adheres to the dynamic diameter $D$.
earliest subsequence in $\sigma$ with a maximal common root 
$R$ in at least $x+1$ consecutive rounds actually has a $\ECS(x+1)$-common root.
\end{definition}

Combining these two definitions results in the following strong version of our
stabilizing message adversary.

\begin{definition} \label{def:AltEStableMA}
The strong stabilizing \MA{} $\AltEStable = \AltSafety(D) + \AltLiveness(D)$ 
contains all graph sequences in
$\AltSafety(D) \, \cap \, \AltLiveness(D)$
that guarantee a dynamic diameter of $D$.
\end{definition}

Note carefully that the very first $\ECS(D+1)$-common root $R'$ occurring
in $\sigma \in \AltEStable$ need \emph{not} be the $\ECS(D+1)$-common root $R$
guaranteed by \cref{def:alt-liveness}.

The following \cref{lem:Adv-comparison} shows that the 
\MA{} $\AltEStable$ is indeed weaker than $\EStable$.
This is not only favorable in terms of model coverage,
but also ensures that an algorithm designed for 
$\AltEStable$ works under $\EStable$ as well.

\begin{lemma}
  $\EStable \subseteq \AltEStable$
  \label{lem:Adv-comparison}
\end{lemma}

\begin{proof}
  Pick any graph sequence $\sigma \in \EStable$.
  Since $\sigma \in \Liveness$, there exists a round $\rSR \geq \rGST$
  such that $\Seqr{\rSR}{\infty}$ is $R$-single-rooted.
  But then $\Seqr{\rSR}{\rSR+D}$ is also $R$-single-rooted and
  there is a set of $D$ additional communication graphs
  $S = \set{\G^{\rSR+D+1}, \ldots, \G^{\rSR+2D}}$
  such that every $\Gr \in S$ is also $R$-single-rooted.
  Hence, $\sigma$ satisfies $\AltLiveness(D)$.

  Furthermore, $\sigma$ satisfies $\Safety(D)$.
  Thus, for the first sequence $\Seqr{a}{a+D}$ with common root $R$,
  $R$ must already be the \FAES-common root and hence 
  $\Seqr{\rSR}{\infty}$ is $R$-single rooted for some $\rSR \geq a$.
  Consequently, $R$ is a $\ECS(x+1)$-common root starting
  at $a$. Hence, $\sigma$ satisfies $\AltSafety(D)$.
\end{proof}

The following \cref{thm:correctness'} shows that the algorithm from \cref{fig:algorithm} also
solves consensus under the stronger \MA{} $\AltEStable$:

\begin{theorem} \label{thm:correctness'}
  For a graph sequence $\sigma \in \AltEStable$,
  let $\G^{r_1}, \ldots, \G^{r_D}$ with $r_1 > \rSR+D$ 
denote the $D$ re-appearances of the $\ECS(D+1)$-common root $R$ 
guaranteed by $\AltLiveness$ according to \cref{def:alt-liveness}.
Then, the algorithm from \cref{fig:algorithm} correctly terminates
by the end of round $\tau=r_D$.
\end{theorem}

\begin{proof} 
The proof of validity in \cref{thm:correctness} is not affected by 
changing the \MA{}.

  For the agreement condition, recall that $\AltSafety(D)$ guarantees that the first sequence
  $\SeqrI$
  with common root $R$ in $D+1$ consecutive rounds has a $\ECS(D+1)$-common root.
  Hence, we can again apply \cref{lem:agreement} to prove that the algorithm 
satisfies agreement.

  For the termination condition, recall that for any sequence $\sigma \in \AltLiveness(D)$
  it is guaranteed that there exists
  some round $\rSR$ s.t.\ $\Seqr{\rSR}{\rSR+D}$ is $R$-single-rooted.
  Furthermore, $\sigma$ contains at least $D$ not necessarily subsequent
  $R$-single rooted communication graphs after $\rSR+D$.
  The latter implies, by \cref{def:dynamic-diameter}, that $\setinfl{R}{\rSR+D}{p}{\tau}$
  for every process $p \in \Pi$.
  Hence, we can again apply \cref{lem:termination}, which shows
  that the algorithm indeed terminates by round $\tau$.
\end{proof}

By contrast, the algorithm from \cite{SWSBR15:NETYS} does not work
under $\AltEStable$.
Under an appropriate adversary, this algorithm ensures graceful degradation
from  consensus to general $k$-set agreement. This does not allow the algorithm
to adapt to the comparably shorter and weaker stability periods of 
$\AltEStable$, however.
In more detail, $\mathtt{VSRC}(n, 4D)$ requires a four times longer period of
consecutive stability than $\AltLiveness(D)$.
The adversarial restriction $\mathtt{MAJINF}(k)$ that enables 
$k$-agreement under partitions in \cite{SWSBR15:NETYS} for $k>1$, on
the other hand, is very weak and thus requires quite involved 
algorithmic solutions. Nevertheless, despite its weakness,
it is not comparable to $\AltSafety(D)$.

\subsection*{Impossibility results and lower bounds}

The proof of \cref{thm:correctness'} indicates that two things are needed in order to solve
consensus under a message adversary like $\AltEStable$:
There must be some subsequence with a single root component $R$ in at least
$x+1$ rounds, and, for every process in the system, there must be some round $r$
such that $R$ appears in the causal past $\CP{p}{r}{\rGST+x}$ .
Looking more closely at the \MA{} $\AltEStable$, it is hence tempting to
further weaken it by instantiating $\AltSafety(x)$ with some $x >D$ 
and/or $\AltLiveness(x)$ with some $x < D$.
There is, however, a fundamental relation between the $\AltSafety(x)$ and
$\AltLiveness(x)$ conditions:
Weakening one condition requires strengthening the other, and vice-versa.

To further explore this issue, we introduce the message adversary
$\MAD(x,y)$, which consists of the graph sequences
in $\AltSafety(x) \cap \AltLiveness(y)$ that guarantee a dynamic diameter $D$.
The following \cref{thm:alt-adversary-optimality} reveals that solving consensus
requires $y \geq x$.
\iftoggle{TR}{}{The proof was, due to space constraints and its similarity with
  \cref{thm:adversary-tightness}, relegated to \cite{SSW15:TR}.}

\begin{theorem}
%  Let $x \geq D$.
  Solving consensus is impossible under \MA{} $\MAD(x,y)$ for $x > y$.
  \label{thm:alt-adversary-optimality}
\end{theorem}

\iftoggle{TR}{
\begin{proof}
Since the processes have no knowledge of 
$\Pi$ and $\abs{\Pi}$, we can again w.l.o.g.\ assume that $n \geq 4$ and 
$D < n-2$. 

  Assume for a contradiction that some algorithm $\A$ exists that solves
  consensus under $\MAD(x,y)$ for $x > y$, and hence also in the following 
execution $\varepsilon$ with graph sequence $\sigma$:
  Every process starts with input value $0$ and, for the first
  $x \geq y+1$ rounds,
  $\Seqr{1}{x}$ is $R$-single rooted.
  Then, the communication graphs alternate between being
  $R'$-single-rooted and $R$-single-rooted
  for some root $R' \neq R$.
  Additionally, there are two distinct processes $p$ and $q$ that have only
  incoming edges throughout the entire execution $\varepsilon$. The actual
communication graphs outside $R$ are such that
$\sigma$ has a dynamic diameter $D$. 
  
  Since every $\Gr$ in $\sigma$ is single-rooted, the latter is feasible
for $\AltLiveness(y)$, with
  $\rGST=1$ and the communication graphs 
$\G^{x+2}, \G^{x+4}, \ldots, \G^{x+2D}$ where $R$ re-appears $D$ times.
In addition, as $\sigma$ does not contain any root component that is
common in more than $x$ rounds, it trivially satisfies $\AltSafety(x)$ as
well. By the assumed 
correctness of $\A$ under $\MAD(x,y)$, there is hence some round $\tau$ by which every process
  must have terminated correctly.

  Now consider the following execution $\varepsilon'$, with graph sequence $\sigma'$:
  Each process of $\Pi \setminus \set{p, q}$ starts with input value
  $0$, while $p$ and $q$ start with $1$.
  For every $\Gr$ of $\Seqr{1}{\tau}$ in $\sigma'$, the induced subgraph of
  $\Pi \setminus \set{p, q}$ is the same as in $\sigma$.
  By contrast, the processes $p$ and $q$ are now connected only with each other:
  There is an edge $(q, p)$ in every $\Gr$ and an edge $(p, q)$ in every
  $\Gr$ where $r$ is even.
  Finally, the graph sequence $\Seqr{\tau+1}{\infty}$ forever repeats the
  star-graph $\G$, where the center $p$ has no
  in-edges and an out-edge to every other process.

Clearly, $\sigma'$ is feasible for $\AltLiveness(y)$, with $\rGST=\tau+1$
due to the star-graph sequence
$\Seqr{\tau+1}{\infty}$.
Moreover, $\Seqr{\tau+1}{\infty}$ is the only
subsequence of $\sigma'$ with a common root $R$ and a longer consecutive
duration than $x$.
Since $R$ is a $\ECS(x+1)$-common root of $\Seqr{\tau+1}{\infty}$, $\sigma'$
is feasible for $\AltSafety(x)$.
Since also the dynamic diameter $D$ is adhered to in $\sigma'$,
we have thus that
$\sigma'$ is feasible for $\MAD(x,y)$.
  
  Observe that all processes of $\Pi \setminus \set{p, q}$ have the same state
  in both $\varepsilon$ and $\varepsilon'$
  at the end of round $\tau$.
  Hence, all decide $0$ in $\varepsilon'$ as they do in $\varepsilon$.
  For $p$ and $q$, $\varepsilon'$ is indistinguishable from the 
execution $\varepsilon''$, which applies $\sigma'$ to the initial configuration
where every process started with input value $1$. Consequently, $p$ cannot make a 
safe decision in $\varepsilon'$: If it decides 1, it violates agreement w.r.t.\ $\varepsilon$,
if it decides 0, it violates validity w.r.t.\ $\varepsilon''$.
  This contradicts the assumption that $\A$ is a correct consensus algorithm.
\end{proof}

Essentially, the proof of \cref{thm:alt-adversary-optimality} exploited the
observation that the members of a root component $R$ cannot distinguish whether they
belong to the single root component guaranteed by $\AltLiveness(y)$ after $\rGST$,
or to a (possibly non-single) ``spurious'' common root in $y+1$ consecutive 
rounds generated my $\MAD(x,y)$
 before $\rGST$. Note that this is closely related to the
argument used for defending the need to introduce $\Safety(x)$ in \cref{def:safety} (recall
the graphs depicted in \cref{fig:stab-is-not-enough}).}{}

In the light of \cref{thm:alt-adversary-optimality}, $\AltEStable$ is hence the strongest
eventually stabilizing variant of $\MAD(x,y)$ for $x \geq D$ we can hope to find
an algorithm for. Note that it would not be difficult to adopt the 
algorithm introduced in 
\cref{fig:algorithm} to work under $\MAD(x,y)$ for general $y \geq x \geq D$, though.
Answering the question of whether it is possible to solve consensus for $x < D$
is a topic of future research.

Finally, \cref{thm:alt-adversary-tightness} provides a termination
time lower bound for consensus under $\AltEStable$.
The result itself is actually a direct consequence of the fact that
$\EStable \subseteq \AltEStable$
(\cref{lem:Adv-comparison}) and \cref{thm:adversary-tightness}.
\iftoggle{TR}{We now}{In \cite{SSW15:TR}, we} provide a more involved argument showing that
the result holds
even for arbitrary choices of $\rSR$ and $\set{r_1, \ldots, r_D}$.

\begin{theorem}\label{thm:alt-adversary-tightness}
  For a graph sequence $\sigma \in \AltEStable$,
  let $\G^{r_1}, \ldots, \G^{r_D}$ with $r_1 > \rSR+D$ 
denote the $D$ re-appearances of the $\ECS(D+1)$-common root $R$ 
guaranteed by $\AltLiveness$ according to \cref{def:alt-liveness}.
  Then, no correct consensus algorithm under the message adversary $\AltEStable$
  can terminate strictly before round $r_D$.
\end{theorem}

\iftoggle{TR}{

\begin{proof}
  We assume w.l.o.g.\ that $n > 4$ and $D < n-3$.
  Furthermore, we do not let the adversary choose
  $\rSR$ and $\set{r_1, \ldots, r_D}$, which results in an even stronger
  impossibility result.

  First, let us define some communication graphs that we employ later on.
  For any graph $\G$, let $\widetilde{\G}$ denote the subgraph of $\G$ induced
  by $\Pi \setminus \set{p_{n-1}, p_n}$, augmented with the edge
  $(p_{n-1}, p_n)$.
  Let $\overline{\G}$ be the same as $\widetilde{\G}$ except that the direction
  of this edge is reversed.
  In addition, let $\G'$ be a graph where $D+2$ processes of $\Pi \setminus \set{p_{n-1}, p_n}$
  constitute a chain $C$ (actually, a tree), with head $p_1$ and two tails $p_{n-3}, p_{n-2}$,
  where the processes of $\Pi \setminus C$ only have incoming edges.
  Let $\G''$ be the same as $\G'$, except that the direction of all the edges
  in $C$ is reversed and there is an edge $e=(p_{n-3}, p_{n-2})$ in $\G''$.
  Let $\G'''$ be the same as $\G''$ but with reversed direction of this edge $e$.

  For a contradiction, assume that an algorithm $\mathcal{A}$ exists that solves
  consensus in a round $\tau \leq$ $r_D - 1$.
  Then, $\mathcal{A}$ must solve consensus also in the following execution $\varepsilon$:
  Let all processes start with input $0$, and
  construct $\sigma = \SeqInf$ as follows:
  For $r \notin \set{r_1, \ldots, r_D}$ and $1 \leq r < \rSR$ or $r > \rSR+D$,
  if $r$ is even, let $\Gr = \G''$; if $r$ is odd, $\Gr = \G'''$.
  For $\rSR \leq r \leq \rSR+D$ or $r \in \set{r_1, \ldots, r_D}$,
  let $\Gr = \G'$.
  Clearly, $\sigma \in \AltEStable$.
  By validity and the assumptions on $\mathcal{A}$, all processes of $\Pi$ must
  decide $0$ by round $\tau$.

  We now define another execution $\varepsilon'$, where
  all processes in $\Pi \setminus \set{p_{n-1}, p_n}$ start with $0$ and
  $p_{n-1}$ and $p_n$ start with $1$.
  The graph sequence $\sigma'$ of $\varepsilon'$ is the same as
  $\sigma$ until round $\tau$,
  except that every $\Gr$ of $\sigma$ is replaced with
  $\widetilde{\Gr}$ if $r$ is even and $\overline{\Gr}$ if $r$ is odd.
  Moreover, $\G'$ in round $\rSR+D$ is not only replaced with $\widetilde{\G}'$,
  but also augmented with a single edge $(p_2,p_1)$.
  Finally, let the $\Gr$ of $\Seqr{\tau+1}{\infty}$ in $\sigma'$ be a star-graph
  with an out-edge from $p_n$ to every process of $\Pi$.
  Again, note that $\sigma' \in \AltEStable$.

  Observe that, in $\sigma'$, for any round $r<r_D$, it holds that
  $p_1 \not\in \CP{p_{n-2}}{r}{\rSR+D}$.  
  Hence, until round $r$, $\varepsilon$ is indistinguishable for 
  $p_{n-2}$ from the execution $\varepsilon'$.
  In particular, $p_{n-2}$ can not have learned about the existence of the edge 
  $(p_2,p_1)$ in $\G^{\rSR+D}$.
  Therefore, since $p_{n-2}$ decides $0$ in
  round $\tau$ in $\varepsilon$, it does so also in $\varepsilon'$.
  This, however, means that $p_n$ can never make a safe decision in $\varepsilon'$:
  In order to satisfy agreement it should decide $0$.
  However, since $p_n$ never hears from process that had input $0$,
  $\varepsilon'$ is indistinguishable for $p_n$ from an execution $\varepsilon''$,
  which has the same graph sequence $\sigma'$ but where all processes have input $1$.
  In order to satisfy validity, it should decide $1$ in $\varepsilon''$. This provides
the required contradiction.
\end{proof}

\subsection*{More efficient algorithms}

Throughout our paper, we have assumed a full-information protocol
where, every round, a process stores and forwards its entire known state history.
While this is a convenient abstraction for introducing the fundamental concepts 
of our algorithm and a valid assumption for any impossibility result,
it is of course highly unpractical.

We can name two major improvements related to this issue.
For simplicity, we only discuss the graph approximation here and not the
matrix $\lock{p}$ of lock values.
It is not hard to see that arguments for the former extend in a natural way
to the latter.

First, it has already been shown, via the graph approximation algorithm used in
\cite{BRS12:sirocco}, that it is sufficient to store and forward the local graph 
approximation history of each process in order to faithfully approximate 
the communication graph sequence. In round $r$, this requires up to $O(r n^2)$ 
local memory space at every process.

Second, the question arises whether it is indeed necessary to maintain (an
approximation of) the entire communication graph sequence.
In the case of $\EStable$, it is perfectly possible to locally store and forward
only a relatively small part of the graph approximation:
Since the largest possible latency for a process to detect the start of
a single-rooted graph sequence of duration $D+1$ is $D$ rounds, it suffices
to maintain only the last $2D+1$ rounds of the graph approximation history.
This optimization yields a memory complexity of $O(D n^2) = O(n^3)$ by 
\cref{cor:end-to-end}.

In the case of $\AltEStable$, there is a tradeoff between the strength of the
adversary and the memory complexity required by the algorithm.
The principal issue is that if we allow the algorithm to purge 
the graph approximations for all but the last $x$ rounds, then 
the adversary could generate a run with a ``terminating'' 
$\ECS(D+1)$-common root $R$ with $r_D > \rSR+D+x$, recall
\cref{def:alt-liveness}. In this case, process $p \in \Pi$ in round $r_D$ 
would not have its causal past down to round $\rSR+D$ available, which 
is mandatory for detecting $R$.

A straightforward remedy would be an additional restriction to be enforced by
the message adversary, which must ensure $r_D \leq \rST+D+x$
for some given additional parameter $x$. A \MA{} weakened in such a way
would entail a memory complexity of $O(x n^2)$ for our consensus algorithm.
}{}

\section{Conclusion} \label{sec:conclusion}
We introduced an eventually stabilizing message adversary
for consensus in a synchronous dynamic network with directed communication. 
Such a model closely captures the behaviour of a real network with
arbitrarily irregular interconnection topology for a finite 
initial period, before it eventually starts to operate in
a reasonably well-orchestrated manner. 

Our message adversary eventually asserts a single strongly connected component without
incoming edges from outside the component, which consists of the same 
set of processes, with possibly changing interconnection topology, either
forever ($\EStable$) or, in a generalized and stronger variant, for a certain
number of (partly consecutive) rounds ($\AltEStable$).
We established that no deterministic algorithm can terminate earlier
than $2D+1$ rounds after stabilization in some execution under $\EStable$, where $D$
is the dynamic network diameter guaranteed by the message adversary,
and provided a matching algorithm, along with its correctness proof,
that even works under $\AltEStable$.

Part of our future work in this area will be devoted to 
finding even stronger message adversaries for stabilizing
dynamic systems, and to the development of techniques for exploiting 
them algorithmically.
%This will help us design algorithms that work on an even broader spectrum of 
%adversaries.

\begin{comment}
\todo{last chance to brag...}

Open questions:
\begin{compactitem}
\item Solvability for $1 \leq Y \leq D$
\item unbeatable protocol
\item alternative formulations of an eventually stabilizing adversary 
that are not on a per-round-graph or root component basis (product graphs...)
\end{compactitem}
\end{comment}

\iftoggle{TR}{\bibliography{lit_bib/lit}}
{{\footnotesize\bibliography{minilit}}}
\bibliographystyle{abbrv}

\iftoggle{TR}{}{
\newpage
\appendix
%\medskip
\section*{Relating $\EStable$ to $\mathtt{VSRC}(n,4D)+\mathtt{MAJINF}(k)$}
%\medskip

\section*{Proof of \cref{lem:agreement}}
%\medskip

}

\end{document}